\documentclass[aps,11pt,prd,groupedaddress,nofootinbib,notitlepage,eqsecnum,preprintnumbers]{revtex4-2}
\usepackage[utf8]{inputenc}
\usepackage{hyperref}
\usepackage{xcolor}
\usepackage{graphicx}
\usepackage{amsmath,amssymb}
\usepackage{bm}
\usepackage{comment}
\usepackage{tabularx}
\usepackage[shortlabels]{enumitem}
\usepackage{overpic}
\usepackage{ulem}
\newcolumntype{Y}{>{\centering\arraybackslash}X}

\begin{document}
\title{Scalar-induced gravitational wave interpretation of PTA data: 
\\
the role of scalar fluctuation propagation speed
}

\author{\textsc{Shyam Balaji$^{a,b}$}}
    \email{{sbalaji}@{lpthe.jussieu.fr}}
\author{\textsc{Guillem Dom\`enech$^{c}$}}
    \email{{guillem.domenech}@{itp.uni-hannover.de}}
\author{\textsc{Gabriele Franciolini$^{d}$}}
\email{{gabriele.franciolini}@{uniroma1.it}}

\affiliation{$^a$Laboratoire de Physique Th\'{e}orique et Hautes Energies (LPTHE), \\
    UMR 7589 CNRS \& Sorbonne Universit\'{e}, 4 Place Jussieu, F-75252, Paris, France}
\affiliation{$^b$Institut d’Astrophysique de Paris, UMR 7095 CNRS \& Sorbonne Universit\'{e}, 98 bis boulevard Arago, F-75014 Paris, France}
\affiliation{$^c$Institute for Theoretical Physics, Leibniz University Hannover, Appelstraße 2, 30167 Hannover, Germany.}
\affiliation{$^d$ Dipartimento di Fisica, Sapienza Universit\`a di Roma and INFN, Sezione di Roma,
Piazzale Aldo Moro 5, 00185, Rome, Italy}

\begin{abstract}
Pulsar timing arrays gathered evidence of the presence of a gravitational wave background around nHz frequencies. If the gravitational wave background was induced by large and Gaussian primordial fluctuations, they would then produce too many sub-solar mass primordial black holes. We show that if at the time of gravitational wave generation the universe was dominated by a canonical scalar field, with the same equation of state as standard radiation but a higher propagation speed of fluctuations, one can explain the gravitational wave background with a primordial black hole counterpart consistent with observations. 
Lastly, we discuss possible ways to test this model with future gravitational wave detectors.  
\end{abstract}

\maketitle

\section{Introduction \label{sec:intro}}

There has been mounting evidence of the presence of a Gravitational Wave Background (GWB) around nHz frequencies in Pulsar Timing Arrays (PTAs) since 2021 \cite{Goncharov:2021oub,Chen:2021rqp,Antoniadis:2022pcn}. The recent data release and analysis of the NANOGrav collaboration 
\cite{NG15-SGWB,NG15-pulsars}, as well as the EPTA/InPTA \cite{EPTA2-SGWB,EPTA2-pulsars,EPTA2-SMBHB-NP}, PPTA \cite{PPTA3-SGWB,PPTA3-pulsars,PPTA3-SMBHB} and CPTA \cite{CPTA-SGWB}, set an approximate amplitude of the GWB around $\Omega_{\rm GW}h^2\sim 10^{-8}$ at $f\sim 10^{-8} {\rm Hz}$. Assuming a free spectrum, NANOGrav finds that $\Omega_{\rm GW}\propto f^\alpha$ with $\alpha=[1.3,2.4]$ at $1\sigma$. As there is no conclusive evidence pointing towards the nature of the source yet, be it new physics and/or mergers of supermassive black holes, it is interesting to investigate the implications of the possible signal for early universe physics. We will focus on new physics in Gravitational Waves (GWs) induced by large primordial fluctuations \cite{Tomita,Matarrese:1992rp,Matarrese:1993zf,Ananda:2006af,Baumann:2007zm,Saito:2008jc,Saito:2009jt}, commonly known as Scalar Induced Gravitational Waves (SIGWs), see e.g. Refs.~\cite{Domenech:2021ztg,Yuan:2021qgz} for recent reviews.

In the NANOGrav analysis \cite{NG15-NP}, and many subsequent papers \cite{Franciolini:2023pbf,Franciolini:2023wjm,Inomata:2023zup,Cai:2023dls,Wang:2023ost,Liu:2023ymk,Unal:2023srk,Figueroa:2023zhu,Yi:2023mbm,Zhu:2023faa,Firouzjahi:2023lzg,Li:2023qua,You:2023rmn}, it is assumed that SIGWs are generated by large primordial adiabatic fluctuations when the universe is dominated by an adiabatic perfect fluid of relativistic particles (see e.g. Ref.~\cite{Huang:2023chx,Gouttenoire:2023nzr,Depta:2023qst} for supermassive primordial black holes). While this is the simplest assumption and motivated by extrapolating our knowledge from the Cosmic Microwave Background (CMB) and Big Bang Nucleosynthesis (BBN), the temperature (or better the redshift) at which the SIGWs were generated lies in a regime we have not probed yet. For instance, the lower bound from BBN for which the universe must be dominated by standard radiation and have thermalized  is $T>4\,{\rm MeV}$ \cite{Kawasaki:1999na,Kawasaki:2000en,Hannestad:2004px,Hasegawa:2019jsa}. GWs in the frequency range that NANOGrav probes, roughly $10^{-9}-10^{-7}\,{\rm Hz}$, were generated between $40\,{\rm MeV}$ and $3\,{\rm GeV}$ (noting that the QCD phase transition happens around $100\,{\rm MeV}$). Thus, there is some margin where new physics could show up. This could include a different equation state of the universe \cite{Assadullahi:2009nf,Inomata:2019zqy,Inomata:2019ivs,Inomata:2020lmk,Papanikolaou:2020qtd,Domenech:2020ssp,Domenech:2021wkk,Dalianis:2020gup,Hajkarim:2019nbx,Bhattacharya:2019bvk,Domenech:2019quo,Domenech:2020kqm,Dalianis:2020cla,Abe:2020sqb,Witkowski:2021raz}, different propagation speed of fluctuations \cite{Domenech:2021ztg,Balaji:2022dbi} and different initial conditions \cite{Domenech:2021and}. These unique scenarios can be tested with the GWB. 

Here we will consider the possibility that the content of the universe was not an adiabatic perfect fluid, but a perfect fluid with a constant propagation speed of fluctuations. For example, a canonical scalar field rolling down an exponential potential \cite{Lucchin:1984yf} has an arbitrary equation of state $w$ and propagation speed $c_s^2=1$. As the simplest case, we will consider that the perfect fluid has equation of state $w=1/3$ and arbitrary $c_s$. Although we have that $w=1/3$ independent of $c_s$, the different propagation speed will affect the SIGW spectrum shape and amplitude \cite{Balaji:2022dbi} as well as the abundance of the associated Primordial Black Holes (PBHs). We will present the case of general $w$ in a subsequent study.

Even though the NANOGrav data has relatively large errors, the best frequency bins lie in the low frequency band, that is $f\sim 10^{-9}-10^{-8}\,\rm Hz$, and data seem to follow a blue tilted power-law, which is better fitted by the low frequency tail of SIGWs \cite{NG15-NP,Franciolini:2023pbf,Franciolini:2023wjm,Inomata:2023zup,Cai:2023dls,Wang:2023ost,Liu:2023ymk,Unal:2023srk,Figueroa:2023zhu,Yi:2023mbm,Zhu:2023faa,Firouzjahi:2023lzg,Li:2023qua,You:2023rmn}. As argued in Refs.~\cite{Hook:2020phx,Inomata:2023zup}, assuming SIGWs are generated from a peaked primordial spectrum, the low frequency tail of SIGWs is either $f^{3-2|b|}$ for broad peaks\footnote{To be more precise, the peak is broad but decays fast enough for small wavenumbers. If the primordial spectrum decays slower than $k^{3/2}$, then the low frequency tail of the SIGW spectrum does not scale as $f^3$ but scales as the square of the primordial spectrum \cite{Atal:2021jyo,Liu:2020oqe,Xu:2019bdp}.} or $f^{2-2|b|}$ for sharp peaks, where $b=(1-3w)/(1+3w)$ with $w$ the equation of state of the universe \cite{Domenech:2019quo,Domenech:2020kqm}.\footnote{The low frequency tail at $f< 10^{-8}\,{\rm Hz}$ is also affected by the QCD phase transition \cite{Franciolini:2023wjm} in case standard model particles dominates the energy budget when nHz GW modes re-enter the Hubble horizon.} 
As the data suggest $\Omega_{\rm SIGW}\approx f^2$ to be the best fitting shape of the spectrum, one is required to be in the tail of the SGWB. For $w= 1/3$ this could be potentially induced by a sharp peak in the curvature power spectrum. 
Also, because the PTA signal is rather large in amplitude, the peak of the SIGW has to be outside but not too far from the PTA range, as we will see.

Importantly, a large amplitude of primordial fluctuations may lead to an overproduction of PBHs for some of the allowed parameter space \cite{Vaskonen:2020lbd,DeLuca:2020agl,NG15-NP,Dandoy:2023jot,Franciolini:2023pbf,Franciolini:2023wjm,Inomata:2023zup,Cai:2023dls,Wang:2023ost,Liu:2023ymk,Unal:2023srk,Figueroa:2023zhu,Yi:2023mbm,Zhu:2023faa,Firouzjahi:2023lzg,Li:2023qua,You:2023rmn}, although some systematic uncertainties remain in the PBH calculation. See Refs.~\cite{Khlopov:2008qy,Sasaki:2018dmp,Carr:2020gox,Green:2020jor,Escriva:2022duf} for recent reviews on PBHs. One way to remedy this is to invoke large and negative non-Gaussianities of primordial fluctuations, which significantly suppress PBH formation \cite{Franciolini:2023pbf} (see also \cite{Liu:2023ymk,Li:2023qua}). For earlier works on the impact of local non-Gaussianities on the SIGWs see Refs.~\cite{Cai:2018dig,Unal:2018yaa,Atal:2021jyo,Adshead:2021hnm,Abe:2022xur} and Ref.~\cite{Pi:2022ysn} for a recent model which suppresses PBH formation even more than negative local non-Gaussianity.

Another known possibility to suppress PBH formation is to increase the propagation speed of fluctuations. See, e.g.,  Ref.~\cite{Escriva:2020tak} for analytical estimations of PBH formation in general cosmological backgrounds and Ref.~\cite{Escriva:2021aeh} for a recent review. Interestingly, increasing the speed of fluctuations moves the resonant peak of the SIGW to the high frequency region, disappearing entirely when the propagation speed is unity. Thus, there is some hope that in the absence of the resonant peak, the maximum of the SIGW spectrum is allowed by the data to move to low frequencies. This would correspondigly decrease the required amplitude of the primordial spectrum. 
More importantly, modifications to $c_s$ also strongly affect the threshold for BH formation, with the PBH abundance decreasing significantly for larger $c_s$. 
As an interesting bi-product, a higher propagation speed of fluctuations may enhance the high-frequency tail of SIGWs \cite{Balaji:2022dbi}, which has implications for future detectors such as LISA \cite{Barausse:2020rsu,LISACosmologyWorkingGroup:2022jok}, Taiji \cite{Ruan:2018tsw}, DECIGO \cite{Yagi:2011wg,Kawamura:2020pcg} and $\mu$-Ares \cite{Sesana:2019vho}.

One open question of such a model is how one recovers the standard cosmology after the phase of scalar field domination. If we assume that transition happens at low frequency, PTAs would not be sensitive to the transition as the equation of state is that of standard radiation. In any case, the naive expectation is that the transition leading to a change of $c_s$ alone would not significantly affect the amplitude of the SIGW spectrum. For these reasons, we will take an agnostic approach. We will study what the implications are of different propagation speeds for SIGWs and whether the recent PTA data shows preferences for particular values of $c_s$. 
Even more crucially, we will address whether PBH overproduction also constrain SIGW scenarios that attempts to explain the PTA observations when larger values of $c_s$ are considered. This also serves as an exercise for future data analyses and illustrates that not only one could probe the primordial spectrum of fluctuations with SIGWs but the content of the universe at the time of generation as well.

This paper is organized as follows. In \S~\ref{sec:sigw} we review the SIGWs generated in a universe with $w=1/3$ but with general $c_s$. We derive the relations between parameters to explain the NANOGrav amplitude for general values of $c_s$. In \S~\ref{sec:pbh} we compute the PBH abundance using both peak theory and Press-Schecter to show that larger $c_s$ suppresses the PBH abundance even after rescaling the amplitude of the SIGW spectrum to match PTA observations. In \S~\ref{sec:results} we present the results of the Bayesian analysis of the NANOGrav and EPTA data. We discuss possible implications of our SIGW signal for future GW detectors in \S~\ref{sec:future}. We conclude our work in \S~\ref{sec:conclusions}.

\section{Scalar Induced Gravitational Waves \label{sec:sigw}}

Large primordial fluctuations lead to a loud GW signal. The properties of primordial fluctuations, such as the power spectrum and non-Gaussianities, together with the content of the universe after inflation determine the amplitude and shape of the SIGW spectrum. In the case of Gaussian fluctuations, there is a general integral formula for the SIGW spectrum for constant equation of state $w$ and propagation speed of fluctuations $c_s$, which reads
\begin{align}\label{eq:spectraldensitytoday3}
\Omega_{\rm GW,0}h^2=1.62\times 10^{-5}\left(\frac{\Omega_{r,0}h^2}{4.18\times 10^{-5}}\right)\left(\frac{g_\rho(T_{\rm c})}{106.75}\right)\left(\frac{g_{s}(T_{\rm c})}{106.75}\right)^{-4/3}\Omega_{\rm GW,c}\,.
\end{align}
$\Omega_{r,0}h^2$ is the radiation fraction today, $g_\rho(T)$ and $g_s(T)$ are the effective degrees of freedom in energy and entropy density respectively (see Ref.~\cite{Saikawa:2018rcs} for precise numerical fits to $g_\rho(T)$ and $g_s(T)$), and $\Omega_{\rm GW,c}$ is the spectral density of SIGWs evaluated at a time, in the standard radiation dominated universe, when the density fraction of GWs becomes constant. The same applies for $T_c$.

It is important to note that the effective degrees of freedom in the scalar field dominated regime are not the same as in standard radiation. In fact, one should set them to unity. Thus, it is important to stress that the subscript “c” is evaluated when the standard radiation dominated universe is recovered. This means that depending on when the transition occurs, the amplitude of $\Omega_{\rm GW,0}$ might change. We have checked that at most, e.g. if the transition happens right before BBN, the amplitude of $\Omega_{\rm GW,0}$ would be a factor $2$ larger in the scalar field dominated case. This would also help lowering the necessary amplitude of the primordial power spectrum to explain the PTA data as well as the amount of PBH produced. For simplicity and to illustrate the main point, namely the effect of $c_s$, we will not take into account this model dependent effect. Our results can then be thought of as a conservative estimate.

The spectral density of SIGWs for $w=1/3$ and constant $c_s$ is given by \cite{Domenech:2021ztg}
\begin{align}\label{eq:Phgaussianfinal}
\Omega_{\rm GW,c}=\int_0^\infty dv\int_{|1-v|}^{1+v}du\,{\cal T}(u,v,c_s,w=1/3){{\cal P}_{\cal R}(ku)}{{\cal P}_{\cal R}(kv)}\,,
\end{align}
where ${\cal P}_{\cal R}(k)$ is the primordial spectrum of curvature fluctuations, the transfer function is given by
\begin{align}\label{eq:w13}
{\cal T}(u,v,c_s,w=1/3)=&\frac{y^2}{3c_s^4}\left(\frac{4v^2-(1-u^2+v^2)^2}{4u^2v^2}\right)^2\nonumber\\&\times
\left\{\frac{\pi^2}{4}y^2\Theta[c_s(u+v)-1]
+\left(1-\frac{1}{2}y \ln\left|\frac{1+y}{1-y}\right|\right)^2\right\}\,,
\end{align}
and
\begin{align}
y=\frac{u^2+v^2-c_s^{-2}}{2 uv}\,.
\end{align}
These formulas coincide with the ones in Refs. \cite{Espinosa_2018,Kohri:2018awv} in the limit where $c_s^2=1/3$. It is important to note that Eq.~\eqref{eq:w13} is not valid in the limit of $c_s^2\to 0$, which would be similar to the case of dust. In that case, it has been shown that the transition from $c_s^2=0$ to $c_s^2=1/3$ has a strong impact on the predicted GW spectrum \cite{Inomata:2019ivs,Inomata:2019zqy}. Although it requires a careful calculation, we expect that for $c_s^2\lesssim 0.01$ Eq.~\eqref{eq:w13} is no longer accurate. For completeness, we provide a concrete realization of a scalar field model with independent and constant $w$ and $c_s$ in App.~\ref{app:model}.

\begin{figure}
\includegraphics[width=0.49\columnwidth]{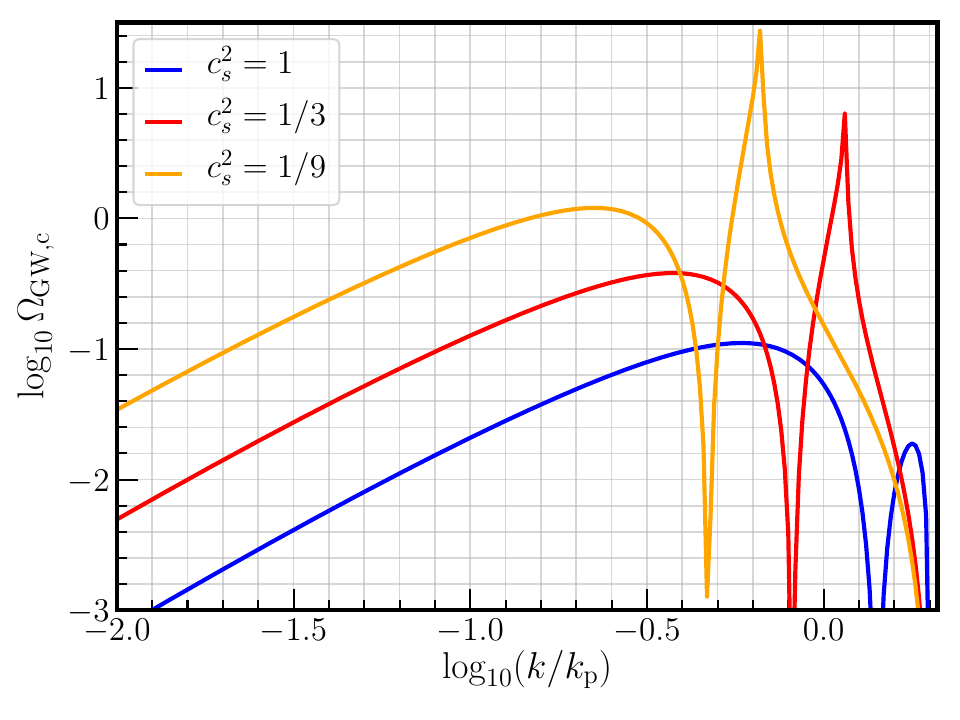}
\includegraphics[width=0.49\columnwidth]{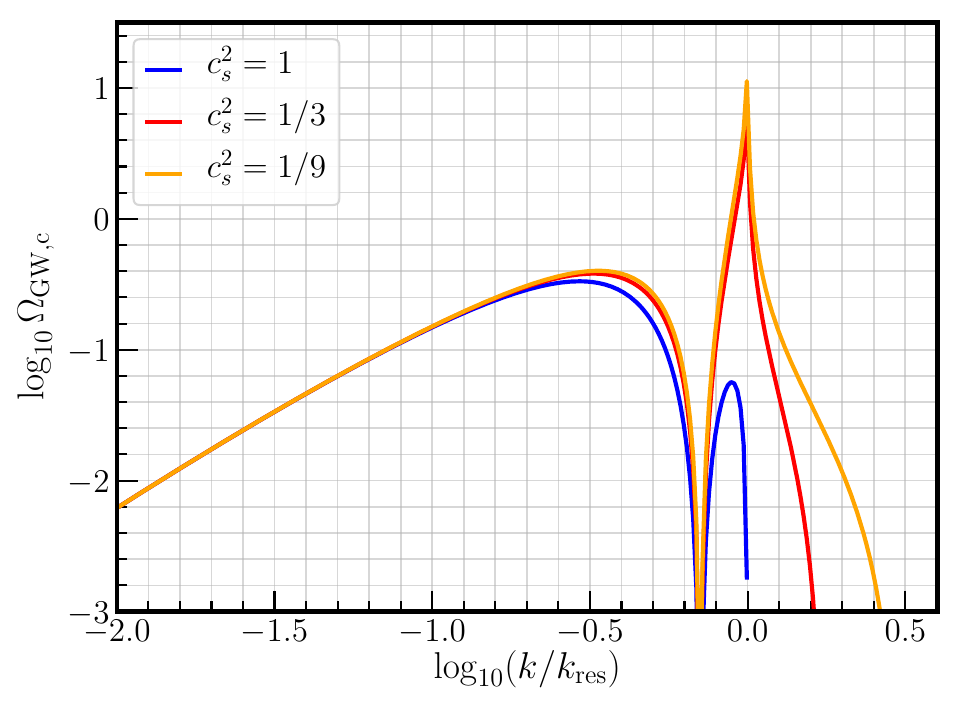}
\caption{Spectral density of scalar induced gravitational waves for $w=1/3$ and constant $c_s$ for a Dirac delta primordial spectrum \eqref{eq:pdirac}. On the left we fix $k_{\rm p}$ and $A_{\cal R}=1$ and vary $c_s^2=1,1/3,1/9$ respectively in blue, red and orange. We note the resonant peak moves to the left and the amplitude increases while decreasing $c_s$. On the right, we fix $k_{\rm bump}$ \eqref{eq:kbump} and require the same amplitude using \eqref{eq:bump}. Then we fix $A_{\cal R}=1$ for $c_s^2=w$. We note how all the curves share the same IR behavior and that the position of the resonant peak coincides. For $c_s=1$ the resonant peak lies precisely at the cut-off. \label{fig:examples}}
\end{figure}

Now, let us specify our modelling of primordial spectrum. For practical purposes, we describe the small scale enhanced spectrum as log-normal peak, namely
\begin{align}\label{eq:PRLN}
{{\cal P}_{\cal R}(k)}=\frac{A_{\cal R}}{\sqrt{2\pi}\Delta}\exp\left[-\frac{\ln^2(k/k_{\rm p})}{2\Delta^2}\right]\,.
\end{align}
We restrict our attention to the case where the spectrum is sharp, i.e. $\Delta<0.2$, since the broad peak case would smear the spectral shape dependence on $c_s$. Thus, in order to obtain simple analytical results, we also consider the limit of $\Delta\to 0$, that is a Dirac delta spectrum given by
\begin{align}\label{eq:pdirac}
{{\cal P}_{\cal R}(k)}=A_{\cal R}\times \delta\left(\ln (k/k_{\rm p})\right)\,.
\end{align}
Interestingly, for $\Delta\lesssim0.2$, we can include the effect of a finite width in the low frequency tail, which is the part relevant for NANOGrav, via \cite{Pi:2020otn}
\begin{align}\label{eq:OmegaDelta}
\Omega^\Delta_{\rm GW,0}h^2={\rm Erf}\left[\frac{1}{\Delta}\sinh^{-1}\frac{k}{2k_{\rm p}}\right]\Omega^\delta_{\rm GW,0}h^2\,.
\end{align}
In Eq.~\eqref{eq:OmegaDelta}, $\Omega^\delta_{\rm GW,0}h^2$ is the result for the Dirac delta case, the analytical expression of which reads
\begin{align}\label{eq:spectraldensitytodaydirac}
\Omega^\delta_{\rm GW,0}h^2\approx 1.8\times 10^{-5}& \frac{A_{\cal R}^2}{12c_s^4}\left(\frac{k}{k_{\rm p}}\right)^2\left(1-\frac{k^2}{4 k_{\rm p}^2}\right)^2\Theta(2k_{\rm p}-k) \nonumber\\&
\times y_{\rm p}^2  \left(\pi ^2 y_{\rm p}^2 \theta \left(1-y_{\rm p}^2\right)+\left(2-y_{\rm p} \log \left(\left|
   \frac{1+y_{\rm p}}{1-y_{\rm p}}\right| \right)\right)^2\right)\,,
\end{align}
where for compactness we defined
\begin{align}
y_{\rm p}=1-\frac{k^2}{2c_s^2k_{\rm p}^2}\,,
\end{align}
and we assumed that $k_{\rm p}\sim 10^{8}\,{\rm Mpc}^{-1}$ to fix the numerical value of the prefactor. We also explicitly write down here the position of the resonant peak, namely
\begin{align}
 k_{\rm res}=2c_sk_{\rm p}\,.
 \end{align}
  We show on the left and right plots of Fig.~\ref{fig:examples} the spectral density for $c^2_s=\{1,1/3,1/9\}$ after fixing $k_{\rm p}$ and $k_{\rm res}$ respectively. 
  On the right plot, we further rescaled the amplitude of the primordial spectrum $A_{\cal R}$ such that the SIGW spectral densities has the same amplitude at low frequencies.

As we have argued, PTA data prefers blue tilted spectra, which may be associated to the low frequency tail of the SIGW background. So, let us derive the conditions such that the amplitude at the tail is the same regardless of $c_s$. This gives a relation between $k_{\rm p}$ and $A_{\cal R}$ in terms of $c_s$. We start by deriving the asymptotic form of the SIGW spectral density \eqref{eq:spectraldensitytodaydirac} for $k\ll k_{\rm p}$, which gives
\begin{align}\label{eq:IRshape}
\Omega_{\rm GW,c}(k\ll k_{\rm p})\approx \frac{1}{12}\left(\frac{A_{\cal R}}{c_s^2}\frac{k}{k_{\rm p}}\right)^2\left(\pi^2+4\left(1+\ln\left[\frac{k}{2c_sk_{\rm p}}\right]\right)^2\right)\,.
\end{align}
As is clear from $\eqref{eq:IRshape}$ there is a degeneracy in the IR tail where, if we neglect the logarithmic correction,
\begin{align}
\frac{A_{\cal R}}{c_s^2}\frac{k}{k_{\rm p}}=\rm constant\,.
\end{align}
Given the upper bound on the amplitude of perturbations induced by PBH overproduction (that we will derive in details in the following), one obtains also a constrain on the maximum scale $k_{\rm p}$ that could be compatible with PTA observations. 

We will find that the data seem to prefer that the amplitude of the smooth bump in the SIGW spectrum, i.e. the maximum that is not the resonant peak (see Fig.~\ref{fig:examples}), is similar for all $c_s$. The position of the “low frequency” maximum is located at
\begin{align}\label{eq:kbump}
k_{\rm bump}= c_s \frac{k_{\rm p}}{\sqrt{2}}.
\end{align}
For reference, the destructive interference minimum where $\Omega_{\rm GW,c}=0$ is located at $k_{\rm des}= \sqrt{2}c_s k_{\rm p}$. The amplitude of the SIGW spectrum at the “low frequency” maximum is given by
\begin{align}\label{eq:bump}
\Omega^{\rm bump}_{\rm GW,c}=\frac{3  }{32 c_s^2}\left(1-\frac{c_s^2}{8}\right)^2\left(\frac{9 \pi ^2}{64}+\left(1-\frac{3
   \ln 7}{8}\right)^2\right)\approx 0.14 A^2_{\cal R}\times\frac{(1-0.12 c_s^2)^2}{c_s^2}\,.
\end{align}
From Eq.~\eqref{eq:bump}, we obtain that for a given $c_s$ the amplitude of the SIGW spectrum at the bump is related to the case of $c_s=\sqrt{w}=1/\sqrt{3}$ via
\begin{align}
\frac{\Omega_{\rm GW,0}h^2(c_{s})}{\Omega_{\rm GW,0}h^2(c_{s}=\sqrt{w})}\propto  \frac{A^2_{{\cal R},c_s}}{A^2_{{\cal R},\sqrt{w}}}\frac{{w}}{c_s^2}\,,
\end{align}
where the subscript $c_s$ and $\sqrt{w}$ respectively refers to a given parameter. For example, $A_{{\cal R},\sqrt{w}}$ refers to $A_{\cal R}$ in the case when $c_s=\sqrt{w}$. We also need to require that the position of the bump \eqref{eq:kbump} is the same for different $c_s$. With this information, we find that the requirement that the low frequency tail and the peak are similar for all models is given by
\begin{align}\label{eq:relations}
A_{{\cal R},c_s}\approx c_s/\sqrt{w}\times A_{{\cal R},\sqrt{w}}
\quad {\rm and} \quad 
k_{{\rm p},c_s} \approx c_s/\sqrt{w}\times k_{{\rm p},\sqrt{w}}\,.
\end{align}
We show the low frequency tail of all spectra coincides on the right plot of Fig.~\ref{fig:examples}. This roughly agrees with the degeneracy we later find in the posteriors of the Bayesian inference on the  NANOGrav and EPTA data, namely a degeneracy along $c^{-1}_s\times A_{\cal R}={\rm constant}$ and $k_{\rm p}\times c_s={\rm constant}$. We now proceed to investigate the effect of $c_s$ on PBH formation.

\section{Primordial Black Holes \label{sec:pbh}}
As we are considering sharp log-normal peaks for the SIGWs, we focus to a good approximation on the case of Dirac delta spectrum \eqref{eq:pdirac} for the PBH counterpart. For definiteness, we will use log-normal \eqref{eq:PRLN} with $\Delta=0.1$ in our numerical calculations.

In order to gain an intuition on the order of magnitude, the typical PBH mass can be estimated to be \cite{Sasaki:2018dmp}
\begin{align}
M_{\rm PBH}=4\pi\epsilon\frac{M_{\rm pl}^2}{H_{\rm p}}\,,
\end{align}
where $M_{\rm pl}$ is the reduced Planck mass, $H_{\rm p}$ is the Hubble parameter when the mode $k_{\rm p}$ enters the Hubble radius ($k_{\rm p}=a_{\rm p}H_{\rm p}$) and $\epsilon$ is the fraction of the Hubble volume that goes into the PBH and has to be determined numerically.
Assuming standard cosmological evolution \cite{Planck:2018vyg}, one has
\begin{align}
M_{\rm PBH,f}\approx 6.7\times 10^{-4}M_\odot\left(\frac{k_{\rm p}}{10^8\,{\rm Mpc}^{-1}}\right)^{-2}\left(\frac{\epsilon}{0.3}\right)\,,
\end{align}
where $M_\odot\approx 2\times 10^{33}\,{\rm g}$ is a solar mass. Notice that the precise value of $\epsilon$ is derived adopting the critical collapse formula, informed wih the results of numerical simulations (see e.g. \cite{Musco:2008hv} and more details in App.~\ref{app:abundance}).
The fraction of PBHs as dark matter can be estiamted to be 
\begin{align}
f_{\rm PBH}\approx 1.4\times 10^{10}\,\beta\,\left(\frac{k_{\rm p}}{10^8\,{\rm Mpc}^{-1}}\right)\left(\frac{g_\rho(T_{\rm p})}{10.75}\right)^{3/4}\left(\frac{g_s(T_{\rm p})}{10.75}\right)^{-1}\,,
\end{align}
where $\beta$ is the energy density fraction of PBHs at formation and $T_{\rm p}$ is the temperature at $H_{\rm p}$. Note that in the case of scalar field domination we should replace $T_{\rm p}$ by $T_{\rm c}$ but it is a minor effect which we neglect.

\begin{figure}
\includegraphics[width=0.5\columnwidth]{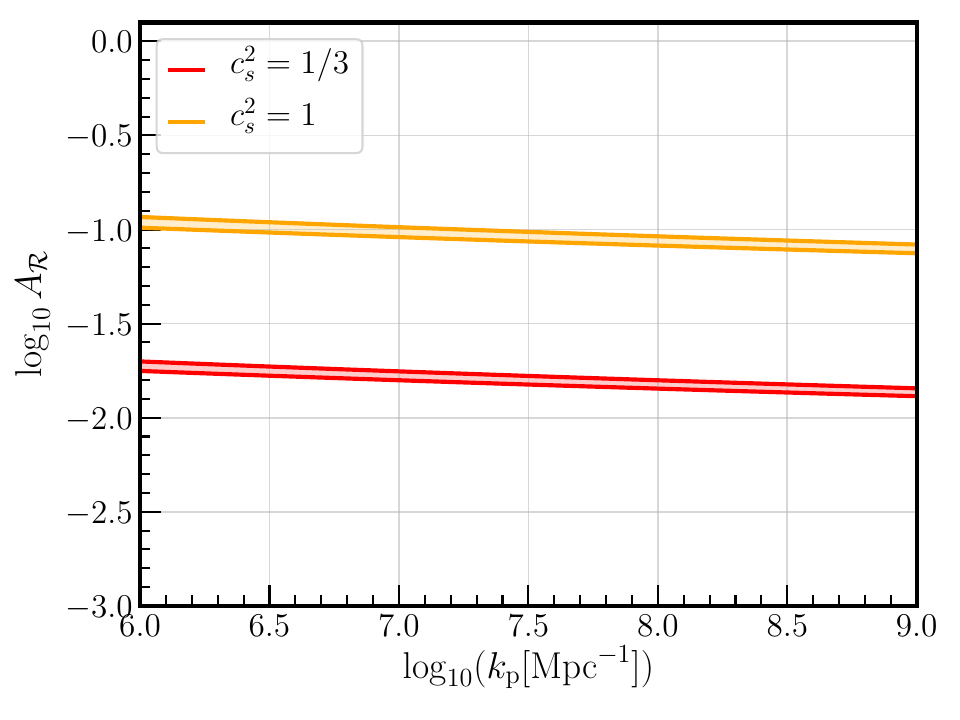}
\caption{Amplitude of primordial spectrum vs peak wavenumber. Solid lines indicate the required amplitude such that $f_{\rm PBH}=1$ using peak formalism (lower line) and Press-Schecter (upper line) with a top-hat window function. To draw the lines, we considered a log-normal peak with $\Delta=1$. See App.~\ref{app:abundance} for numerical details. We show that increasing the propagation speed of fluctuations $c_s$ demands a higher value of $A_{\cal R}$ to form the same amount of PBHs. \label{fig:fpbh}}
\end{figure}

We compute the PBH abundance in detail taking into account the critical collapse and non-linearities as in Refs.~\cite{Young:2019yug,DeLuca:2019qsy,Gow:2020bzo,Young:2022phe,Ferrante:2022mui}.
We perform the computation using both Peak Theory and Press-Schecter, in order to account for the uncertainties related to the choice of method, which results on slightly different prediction for the PBH abundance \cite{Young:2014ana}.
We provide all the formulas and parameters used in App.~\ref{app:abundance} and show the numerical results in Fig.~\ref{fig:fpbh}. 

To understand the effect of a change in $c_s$, it is instructive to use a simplified Press-Schecter approach without the inclusion of non-linearities and critical collapse, where we have that \cite{Sasaki:2018dmp}
\begin{align}
\beta\approx \frac{\sigma_{\rm PBH}}{\sqrt{2\pi}{\delta_{\rm th}}}\exp\left[-\frac{\delta_{\rm th}^2}{2\sigma^2_{\rm PBH}}\right]\,,
\end{align}
where $\delta_{\rm th}$ is the threshold for critical collapse, which we take to be $\delta_{\rm th}=0.59$ \cite{Musco:2020jjb}, and
\begin{align}
\sigma_{\rm PBH}^2  = \frac{16}{81}\int_0^\infty \frac{\mathrm{d}k}{k}(k r_m)^4 {W}^2(kr_m) \mathcal{P}_{\cal R}(k),
\end{align}
with $ {W}^2(kr_m)$ is the window function. We use the real space top-hat window function, which is consistent with the smoothing adopted to determine the threshold for collapse in \cite{Musco:2020jjb}, and given in App.~\ref{app:abundance}.

At this simple level, the only effect of $c_s$ is to increase or decrease $\delta_{\rm th}$. The naive expectation is that increasing $c_s$ would raise the threshold. The actual effect of $c_s$ on  $\delta_{\rm th}$ for a fixed $w$ has not been studied with numerical simulations, but for $c_s^2=1$ we follow Carr’s estimate \cite{Carr:1975qj}, that on constant Hubble slices leads to \cite{Domenech:2020ers}
\begin{align}
\delta_{\rm th}=\frac{3(1+w)}{5+3w}c_s^2=\frac{3(1+w)}{5+3w}=\frac{2}{3}\,.
\end{align}
This corresponds to the maximum value for the threshold one can get by saturating the compaction function. However, choosing $\delta_{\rm th}=2/3$ exactly leaves no room for type-I PBH formation in the critical collapse formulation \cite{Musco:2020jjb}. We therefore take $\delta_{\rm th}(c_s=1)\approx0.66$.\footnote{We also checked that slightly lowering the threshold to $0.65$ does not affect our results in any relevant way, which shows the PBH suppression is not due to the threshold being close to the edge of type-I parameter space.} As this only represents an estimate for $c_s=1$, further numerical investigations are required to determine the precise value of the threshold in these scenarios. We note that the case of generic $c_s^2=w$ has been studied in Ref.~\cite{Musco:2012au,Escriva:2020tak} but it is not possible to extrapolate to general $c_s$ from their analysis.

As the shape of smoothed power spectrum remains the same when varying the equation of state, while only its amplitude varies, we can compare the PBH abundance in the general case of $c_s$ with the well studied case of $c_s=\sqrt{w}$. By doing so, we obtain
\begin{align}
\frac{f_{\rm PBH}^{\rm PS}(c_s)}{f_{\rm PBH}^{\rm PS}(c_s=\sqrt{w})}\approx \frac{\delta_{{\rm th},\sqrt{w}}}{\delta_{{\rm th},c_s}}\frac{k_{{\rm p},c_s}}{k_{{\rm p},\sqrt{w}}}\sqrt{\frac{A_{{\cal R},c_s}}{A_{{\cal R},\sqrt{w}}}}e^{-\frac{{\delta^2_{{\rm th},\sqrt{w}}}}{2\sigma^2_{{\rm PBH},\sqrt{w}}}\left(\frac{A_{{\cal R},\sqrt{w}}}{A_{{\cal R},c_s}}\frac{\delta^2_{{\rm th},c_s}}{\delta^2_{{\rm th},\sqrt{w}}}-1\right)}\,.
\end{align}
Using the relations we found in the previous section, namely Eq.~\eqref{eq:relations}, if we require that the SIGW spectrum computed with different values of $c_s$ provides the explanation for the NANOGrav signal, i.e. the IR tail has its amplitude fixed by requiring consistency with observations irrespectively of $c_s$, the abundance of the PBH counterpart is modified as 
\begin{align}\label{eq:comparison}
\frac{f_{\rm PBH}^{\rm PS}(c_s)}{f_{\rm PBH}^{\rm PS}(c_s=\sqrt{w})}\approx \sqrt{\frac{c_s}{\sqrt{w}}}\frac{\delta_{{\rm th},\sqrt{w}}}{\delta_{{\rm th},c_s}}e^{-\frac{{\delta^2_{{\rm th},\sqrt{w}}}}{2\sigma^2_{{\rm PBH},\sqrt{w}}}\left(\frac{\sqrt{w}}{c_s}\frac{\delta^2_{{\rm th},c_s}}{\delta^2_{{\rm th},\sqrt{w}}}-1\right)}\,.
\end{align}
We therefore conclude that the abundance is suppressed with respect to $c_s=\sqrt{w}$, that is $f_{\rm PBH}^{\rm PS}(c_s)<f_{\rm PBH}^{\rm PS}(c_s=\sqrt{w})$, when
\begin{align}\label{eq:inequalitydelta}
{\delta_{{\rm th},c_s}}>
\left (\frac{c_s}{\sqrt{w}} \right )^{1/2}\times {\delta_{{\rm th},\sqrt{w}}}\,,
\end{align}
because of the exponential dependence in Eq.~\eqref{eq:comparison}.

The inequality \eqref{eq:inequalitydelta} saturates for $\delta_{{\rm th},c_s}=\sqrt{c_s}$.  If we use Carr’s estimate, namely $\delta_{{\rm th},c_s}\propto c_s^2$, we conclude that PBH abundance is always suppressed for larger propagation speed. But Carr‘s value may not a very good estimate in general. For practical purposes, let us assume that there is a general power-law dependence on $c_s$, i.e. $\delta_{{\rm th},c_s}\propto c_s^{n}$. In that case, we find that the inequality \eqref{eq:inequalitydelta} yields
\begin{align}
\left(\frac{c_s}{\sqrt{w}}\right)^{2n-1}>1\,.
\end{align}
For $n>1/2$, the PBH abundance is suppressed if $c_s>\sqrt{w}$. This is the natural expectation: although the amplitude of the primordial spectrum is also larger for larger $c_s$, if we require the same IR tail amplitude of the SIGWs \eqref{eq:relations}, PBH formation is exponentially sensitive to any change. For $n=1/2$ there is no change in the abundance of PBHs. The change in amplitude is exactly compensated by the change in threshold. For $n<1/2$ then PBH abundance is instead suppressed if $c_s<\sqrt{w}$.

In our numerical estimates, we find that $n>1/2$ and PBH abundance is more suppressed with larger $c_s$. In Fig.~\ref{fig:fpbh} we show the values of $A_{\cal R}$ such that $f_{\rm PBH}=1$ for $c_s^2=1/3$ and $c_s^2=1$. It can be seen that the change in the required $A_{\cal R}$ is almost one order of magnitude: from $A_{\cal R}\sim 10^{-2}$ for $c_s^2=1/3$ to $A_{\cal R}\sim 10^{-1}$ for $c_s^2=1$. Thus, the change in $f_{\rm PBH}$ for $c_s^2=1$ largely accommodates the slightly larger amplitude needed yield the same amplitude of the GW spectrum at low frequencies, which is a factor $\sqrt{3}$.

\section{Results \label{sec:results}}

\begin{figure}[t!]
\includegraphics[width=0.67\columnwidth]{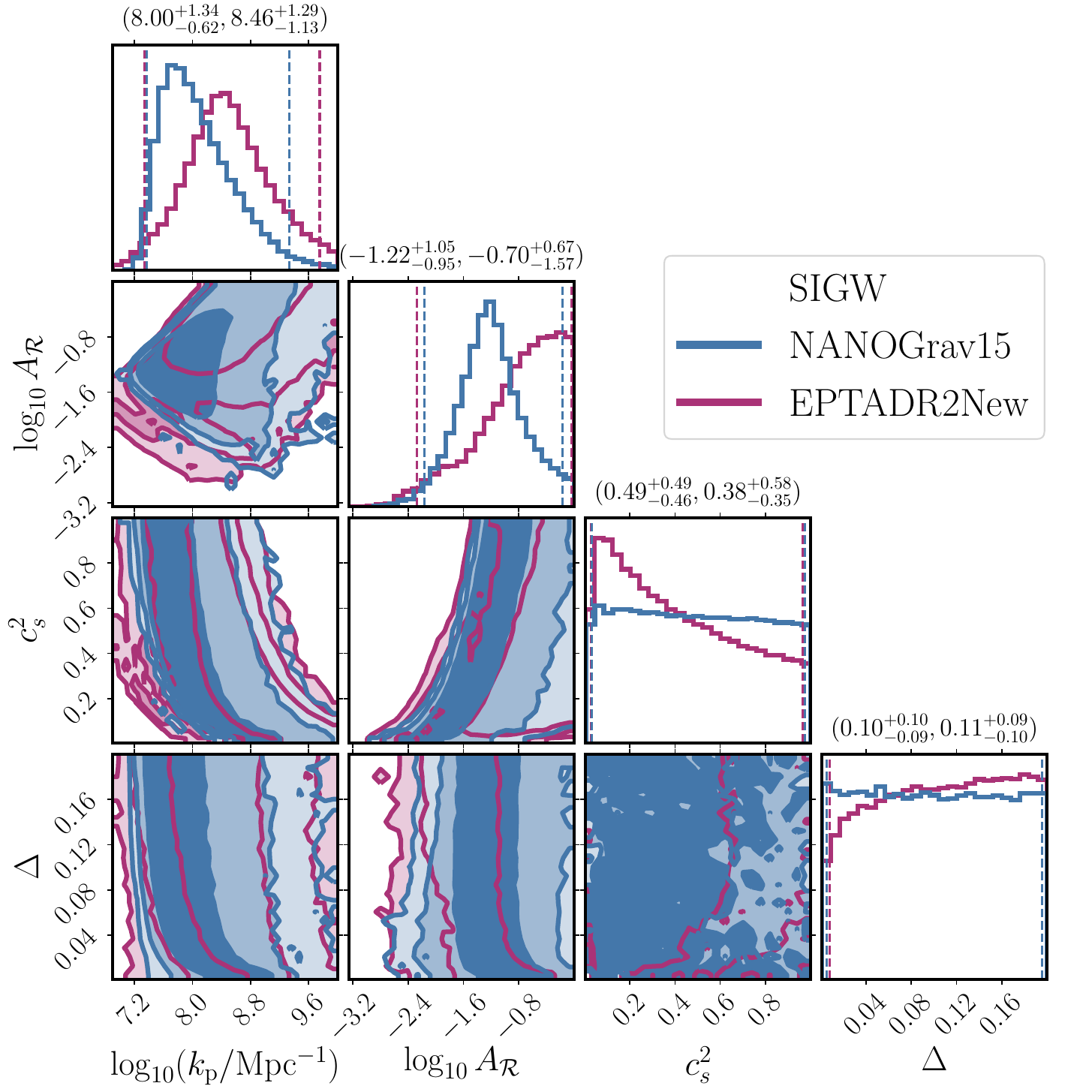}
\caption{ Posterior distributions for NANOGrav (blue) and EPTA (purple) for SIGWs generated during a phase with $w=1/3$ and free $c_s$. Note the degeneracies in the parameters ($c_s$, $k_{\rm p}$) and ($c_s$, $A_{\cal R}$), which agree with the expectations of \eqref{eq:relations}.
\label{fig:fitPTA}}
\end{figure}

We perform a Bayesian inference on the NANOGrav~\cite{NG15-SGWB} and EPTA~\cite{EPTA2-pulsars} datasets, using their posteriors distribution for  $\Omega_{\rm GW}$ in the lowest 14 and 9 frequency bins, respectively, following the choices made by the collaborations.
We adopt our templates \eqref{eq:OmegaDelta} and \eqref{eq:spectraldensitytodaydirac}, which are characterised by the free parameters in the model: $A_{\cal R}$, $k_{\rm p}$, $c_s$ and $\Delta$  
(the corresponding priors adopted in the analysis can be found in Tab.~\ref{tab:priors} in App.~\ref{app:fixedcs}). 
The posteriors distributions are shown in Fig.~\ref{fig:fitPTA}.

We now discuss the results and implications in more detail.
From Fig.~\ref{fig:fitPTA} we first notice that the data is not very sensitive to the value of $\Delta$ as the posterior is essentially flat. This is because the PTA SGWB signal is always explained by the IR tail of the SIGW, which is mildly sensitive to the (small, but finite) width of the curvature spectrum.
The same occurs with the value of $c_s$. Even though EPTA data seems to prefer low values for $c_s$, the tilt in the posterior distribution is not significant. 
Focusing on the posteriors for $c^2_s$, $A_{\cal R}$ and $k_{\rm p}$, we also see the degeneracy explained in \S~\ref{sec:sigw}, Eq.~\eqref{eq:relations}. Namely, $k_{\rm p}\propto c_s^{-1}$ and $A_{\cal R}\propto c_s$. Lastly, we find that the $1\sigma$ contours for $A_{\cal R}$ and $k_{\rm p}$ fall around $A_{\cal R}\sim 10^{-1}$ and $k_{\rm p}\sim 10^8\,{\rm Mpc}^{-1}$ ($f_{\rm p}\sim 10^{-7}\,\rm Hz$).

These results confirm our expectations. Lower values of $c_s$ allow for smaller amplitude of the primordial spectrum, ${A}_{\cal R}$, while higher values of $c_s$ require a higher amplitude. Furthermore, the peak of the SIGW spectrum lies to the right of the PTA data, leading to the degeneracy conditions \eqref{eq:relations} requiring that the IR tail of the SIGW spectrum and the low frequency bump to have the same amplitude for any $c_s$.

To discuss the effect of $c_s$ more clearly, we also performed the analysis with fixed $c_s^2=1/3$ and $c_s^2=1$, which we show in App.~\ref{app:fixedcs}, 
Fig.~\ref{fig:fitPTAcomparisons}.

\begin{figure}[!ht]
\includegraphics[width=0.49\columnwidth]{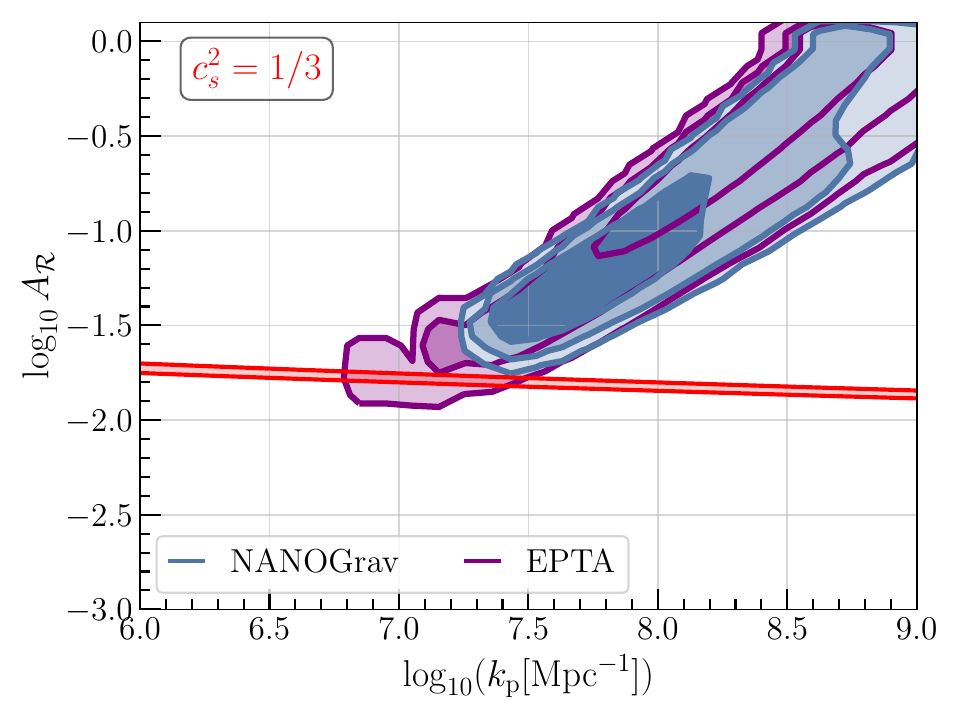}
\includegraphics[width=0.49\columnwidth]{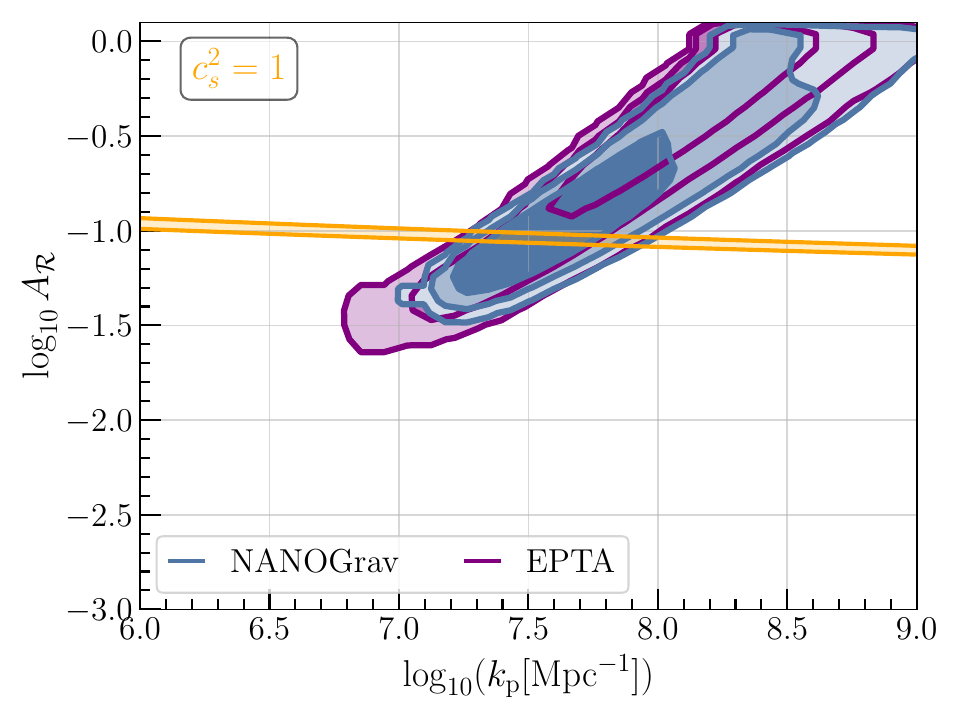}
\caption{Marginalized posterior distributions in the $A_{\cal R}-k_{\rm p}$ plane for fixed $c_s$. In blue and purple we show the $3\sigma$ contours of the NANOGrav and EPTA dataset respectively. In solid, horizontal lines, we also show the requirement on $A_{\cal R}$ such that $f_{\rm PBH}=1$ at a fixed $k_{\rm p}$. To compute $f_{\rm PBH}$, we assumed Gaussian primordial fluctuations and we used peaks formalism (lower line) and threshold statistics (upper line) to illustrate possible uncertainties. Details on the calculations of $f_{\rm PBH}$ can be found in App.~\ref{app:abundance}. On the left figure we present the case of $c_s^2=1/3$. We show how PBH overproduction is about $3\sigma$ tension with the bounds on $A_{\cal R}$ from NANOGrav. On the right figure, we show the case of $c_s^2=1$, where the PBH counterpart is consistent within the $1\sigma$ bounds of NANOGrav and only requires $A_{\cal R}\lesssim10^{-1}$.\label{fig:pbhresults}}
\end{figure}

The most interesting result can be seen in Fig.~\ref{fig:pbhresults}. We show the $3\sigma$ countours in the plane $A_{\cal R}-k_{\rm p}$ for the fixed cases $c_s^2=1/3$ (left plot) and $c_s^2=1$ (right plot). We also show the $f_{\rm PBH}=1$ lines in the figures with some uncertainty depending on whether we use peak theory (bottom line) or Press-Schecter (top line), assuming a log-normal primordial spectrum \eqref{eq:PRLN} with $\Delta=0.1$. We find that while for $c_s^2=1/3$ there is almost a $3\sigma$ tension with the overproduction of PBHs,\footnote{Note that, in general, the value of $A_{\cal R}$ that satisfies $f_{\rm PBH}=1$ may have larger uncertainties depending on the formalism and window functions one uses. So it may be possible to reduce the tension using different choices. That being said, we use the most up-to-date formalism for the calculations.} for $c_s^2=1$ the PTA results are consistent with no overproduction of PBHs inside the $1\sigma$ contours.\footnote{ Notice we do not include the QCD effect on the PBH collapse induced by the corresponding softening of the equation of state \cite{Jedamzik:1996mr,Byrnes:2018clq,Franciolini:2022tfm,Escriva:2022bwe,Musco:2023dak}, as this would require standard model thermal bath to dominate the energy density of the universe, and would only modulate to the standard $c_s^2= 1/3$ case. In any case, this is only has a minor impact on $A_{\cal R}$ at scales below $k_{\rm p} <10^{7}/ {\rm Mpc}$ which is outside the parameter space compatible with PTA observations (see e.g. \cite{Franciolini:2023pbf}). }

From our results, we conclude that if the PTA GWB signal is due to SIGWs, it hints at $c_s^2>1/3$ or non-Gaussian primordial fluctuations (as shown in Refs.~\cite{Franciolini:2023pbf}, see also \cite{Liu:2023ymk,Li:2023qua}), or both. It is interesting to note that while PTAs might not be very sensitive to the value of $c_s$ (or the non-Gaussian parameter $f_{\rm NL}$), the PBH counterpart can reduce the allowed parameter space and prefer certain values of these parameters. Our results also emphasize the potential importance of being agnostic on the content of the universe at the time of wave generation and using the GWB data and PBHs to constrain models of the unexplored early universe.

\begin{figure}[!ht]
\includegraphics[width=0.49\columnwidth]{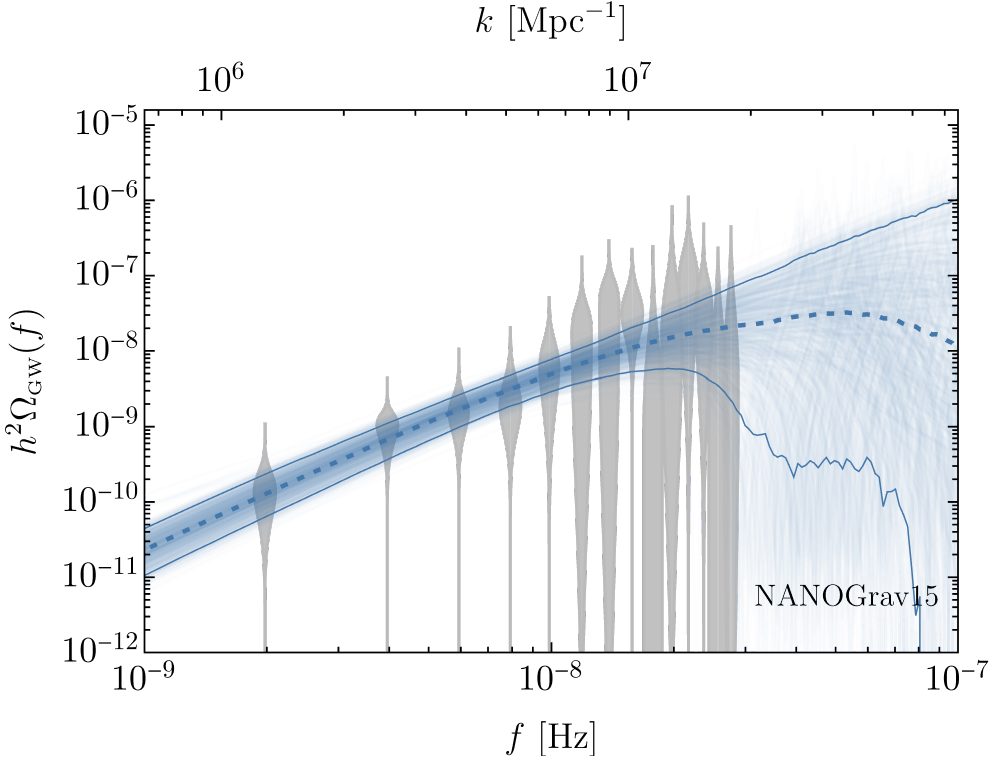}
\includegraphics[width=0.49\columnwidth]{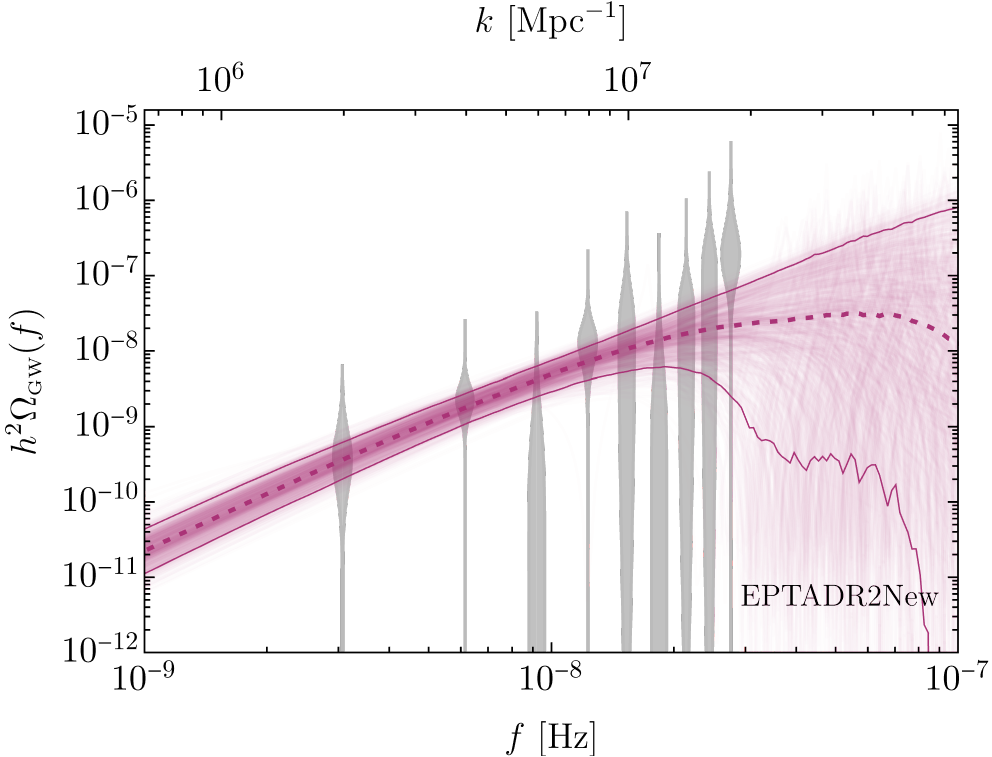}
\caption{ SIGW spectral density for arbitrary $c_s$ parameter (see Eqs.~\eqref{eq:OmegaDelta} and \eqref{eq:spectraldensitytodaydirac}). We respectively show the fit to NANOGrav15 and EPTA data sets on the left and right panels. The colored bands show the 90\% credibility intervals and the gray violins show the posteriors from NANOGrav \cite{NG15-pulsars,NG15-NP}and EPTA \cite{EPTA2-pulsars}.
\label{fig:spectra}}
\end{figure}

 Finally, we also show the SIGW spectrum with the violin plots of NANOGrav and EPTA in Fig.~\ref{fig:spectra} for arbitrary $c_s$, confirming the good agreement between the model discussed in this work and the PTA data (see Fig.~\ref{fig:spectraapp} in App.~\ref{app:fixedcs} for the results fixing $c_s^2=1/3$ and $c_s^2=1$).

\section{Future prospects \label{sec:future}}

In this section, we discuss a characteristic feature of the $c_s^2=1$ case in the high frequency tail of SIGWs which can be tested by future GW detectors.
The appearance of this feature requires the presence of additional power after the log-normal peak \eqref{eq:PRLN} in the curvature power spectrum. 
For instance, in many models, inflation does not end right after the transition giving rise to the peak \eqref{eq:PRLN} and it is not unreasonable to assume that it continues with a second phase of slow-roll. In Ref.~\cite{Balaji:2022dbi}, it has been demonstrated that, in this setting, induced GWs with $c_s^2=1$ present a characteristic $\Omega_{\rm GW}\propto f^{-2}$ scaling at high frequencies before reaching the floor due to the second slow-roll phase. Finding the $f^{-2}$ tail would potentially give further hints towards the  model discussed in this work and would reveal the amplitude of primordial spectrum from the second inflationary stage, even without actually detecting the SIGW low amplitude plateau. For $c_s^{2}<1$ the slope changes to $\Omega_{\rm GW}\propto f^{-4}$, and the sharp drop in the SIGW amplitude hinders any possible detection. For details we refer the reader to \cite{Balaji:2022dbi}. 

For simplicity, we assume that the power spectrum from the second slow-roll stage is a scale invariant spectrum with an enhanced amplitude, which we call ${ A}_{\rm flat}=A_{\cal R}\times A_{\rm rel}$ with $A_{\rm rel}<1$. The actual amplitude of ${ A}_{\rm flat}$ is set by the first slow-roll parameter $\epsilon$ is during the second stage. Note that any contribution from an extrapolation to the almost scale invariant spectrum seen in the CMB \cite{Planck:2018vyg,Akrami:2018odb} is negligible and, therefore, we ignore it. From now on, we will mainly use the parameter $A_{\rm rel}$ for convenience. The total primordial power spectrum is then given by
\begin{align}\label{eq:PR2}
{\cal P}_{\cal R}(k)={ A}_{\cal R}\times \left({\cal P}_{{\cal R},{\rm LN}}(k/k_{\rm p})+{ A}_{\rm rel}{\cal P}_{{\cal R},{\rm flat}}(k/k_{\rm p})\right)\,,
\end{align}
where ${\cal P}_{{\cal R},{\rm LN}}(k/k_{\rm p})$ is the log-normal peak given in Eq.~\eqref{eq:PRLN} with ${ A}_{\cal R}$ factorized and
\begin{align}\label{eq:flat}
 {\cal P}_{\cal R,\rm flat}=\frac{1}{2}\left(1 + \tanh\left[\frac{2}{\Delta}\ln\left(\frac{k}{k_{\rm p}}\right)\right]\right),
\end{align}
where $\Delta$ is the same dimensionless width of the log-normal peak \eqref{eq:PRLN} to avoid adding unnecessary extra parameters.\footnote{The actual shape of ${\cal P}_{\cal R,\rm flat}$ does not matter as long as the step occurs faster or with a similar width than ${\cal P}_{{\cal R},{\rm LN}}$.} In this way ${\cal P}_{\cal R,\rm flat}$ is a smooth step which represents the switch from the first to the second slow-roll level without affecting much the scales of the spike. As in previous sections, we will focus on the case where the log-normal is narrow, i.e. $\Delta<0.2$. We expect the template \eqref{eq:PR2} to be a good approximation to most models where the feature during inflation has sharp transitions from and to the first and second slow-roll phases. For example, see Refs.~\cite{Pi:2017gih,Ando:2018nge,Atal:2018neu,Chen:2019zza,Braglia:2020eai,Ragavendra:2020sop,Fumagalli:2020adf,Ng:2021hll}. For gradual transitions and broad peaks, the distinction we used in Eq.~\eqref{eq:PR2} might not be as accurate. 

Plugging in the ansatz \eqref{eq:PR2} into the SIGW formula \eqref{eq:Phgaussianfinal} we can isolate each contribution as
\begin{align}\label{eq:split}
\Omega_{\rm GW,c}={ A}_{\cal R}^2\times\left(\Omega_{\rm GW,LN}+2{ A}_{\rm rel}\Omega_{\rm GW,\rm cross}+{ A}_{\rm rel}^2\Omega_{\rm GW,flat}\right)\,.
\end{align}
The cross contribution $\Omega_{\rm GW,\rm cross}$ has one ${\cal P}_{{\cal R},{\rm LN}}$ and one ${\cal P}_{{\cal R},{\rm flat}}$ in \eqref{eq:Phgaussianfinal} while the flat contribution $\Omega_{\rm GW,\rm flat}$ has two ${\cal P}_{{\cal R},{\rm flat}}$. ${ A}_{\cal R}^2\Omega_{\rm GW,\rm LN}$ is the same we calculated in previous sections. The additional cross and flat contribution can produce interesting relatively high frequency GW behavior that may be probed with experiments such as LISA, $\mu$Ares and DECIGO.

\begin{figure}[!ht]
\includegraphics[width=0.49\columnwidth]{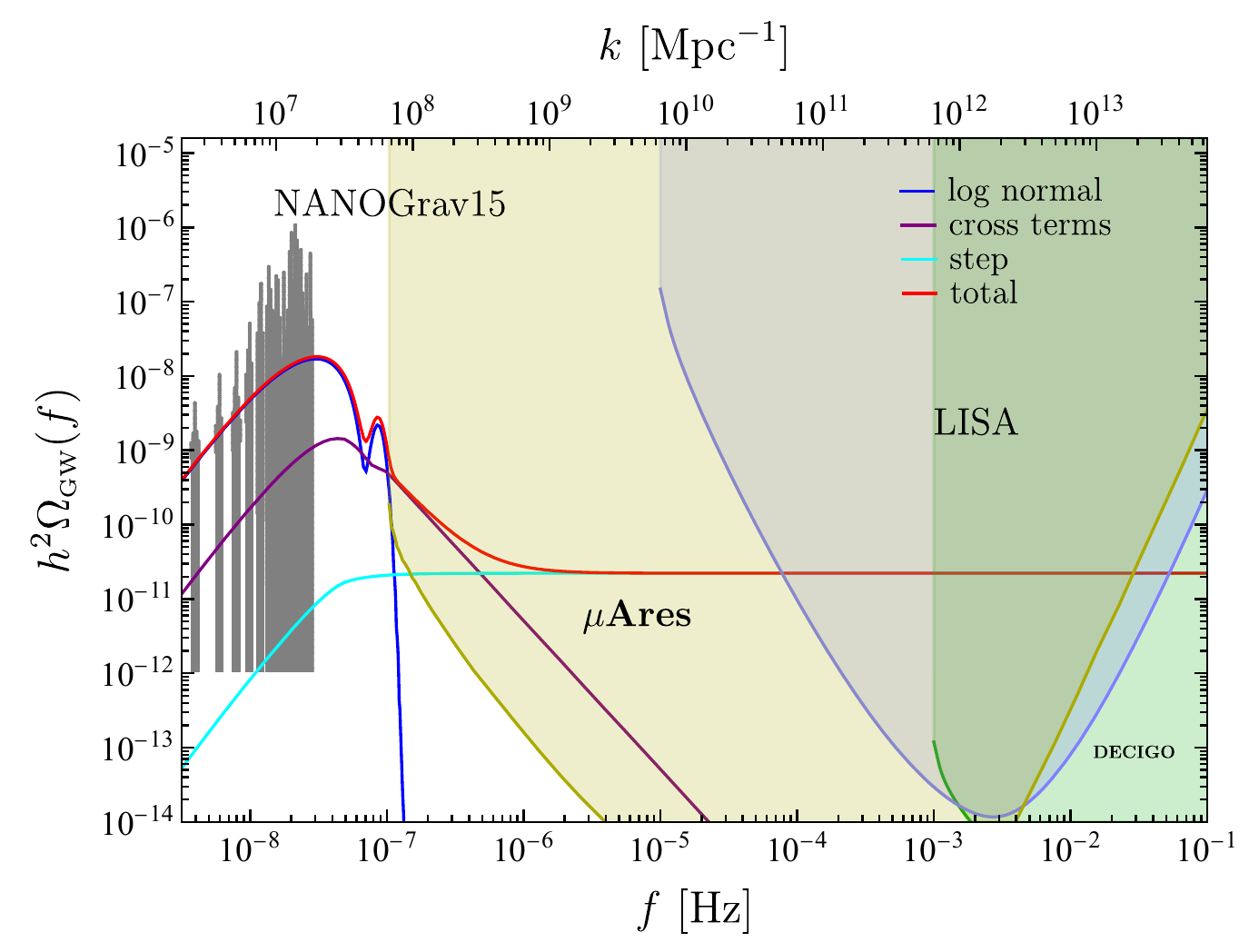}
\includegraphics[width=0.49\columnwidth]{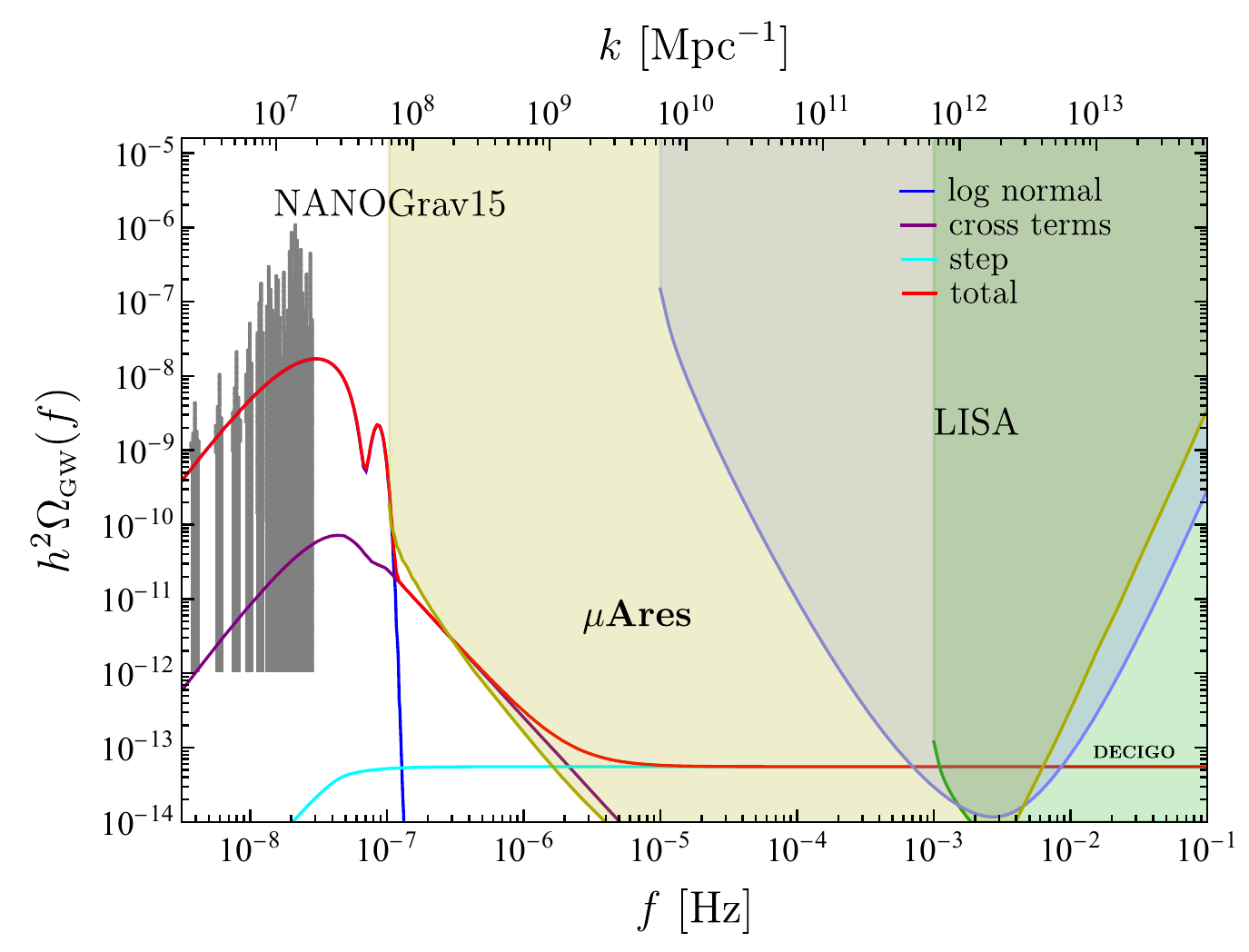}
\includegraphics[width=0.49\columnwidth]{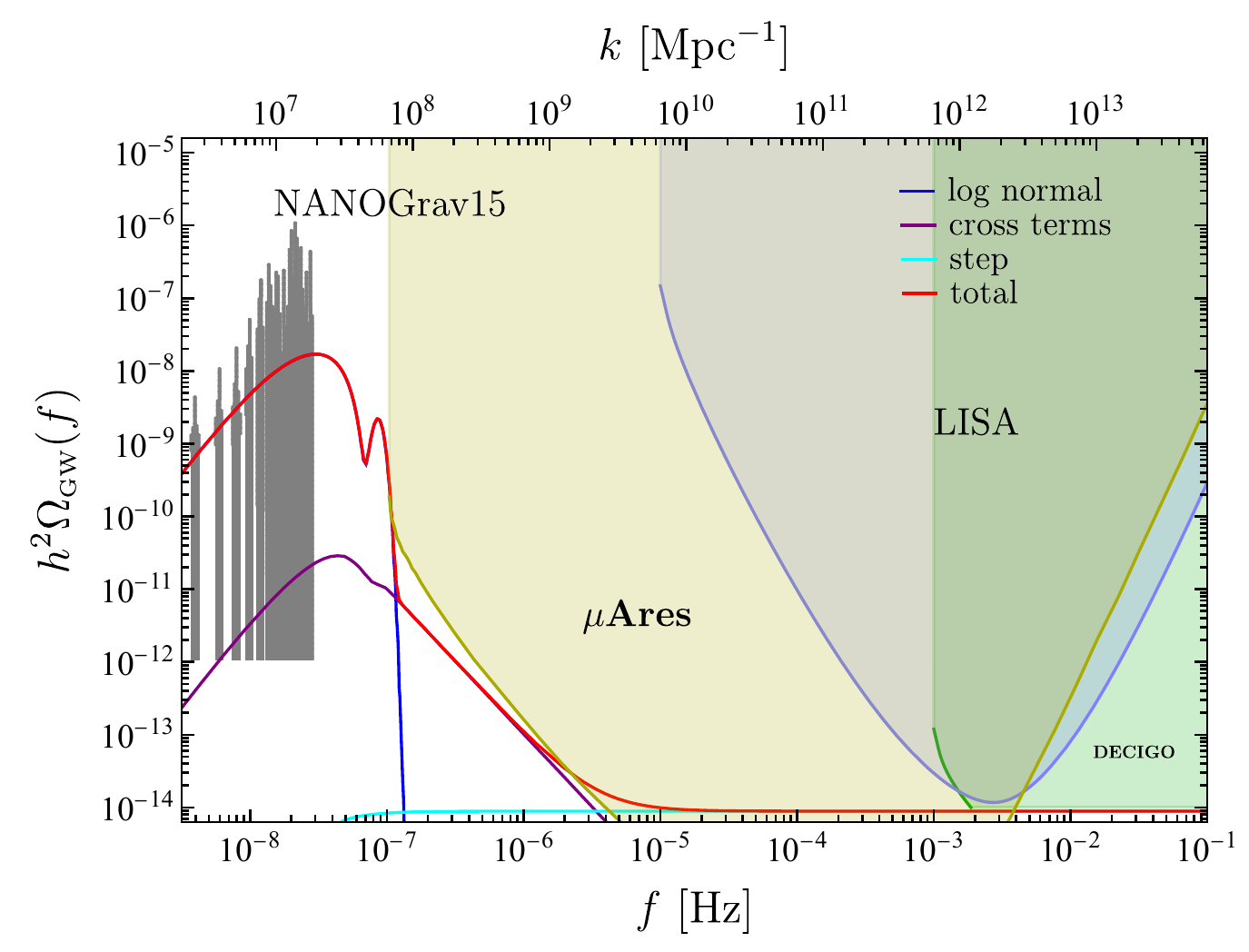}
\caption{We show the gravitational wave spectrum $h^2\Omega_{\textrm{GW}} (f)$ in the present universe as a function of the frequency in Hz for an inflation scenario with a log-normal source term as amplitude $A_{\mathcal{R}}$ and a secondary phase of inflation with amplitude ${A}_{\textrm{flat}}={A}_{\textrm{rel }}{A}_{\cal R}$. We set $A_{\mathcal{R}}=0.1$, $\Delta=0.1$, $k_{\rm p}=3.2\times 10^7\,\textrm{Mpc}^{-1}$ and speed of sound of $c_s=1$. We set $\mathcal{A}_{\textrm{rel }} = 0.01$ (top left), $5\times10^{-4}$ (top right) and $2\times10^{-4}$ (bottom) respectively. We show the relative contributions as log-normal source term (blue), cross terms (purple), step (cyan) and total (red) respectively. For illustration purposes we include the power-law integrated sensitivity curves \cite{Thrane:2013oya} for LISA, $\mu$Ares and DECIGO. \label{fig:crossterms}}
\end{figure}

For concreteness, we consider the posterior distribution of parameters shown in Fig. \ref{fig:fitPTAcomparisons}. We select a point within the 1$\sigma$ region with $\Delta=0.1$, $c_s=1$ and amplitude ${ A}_{\cal R}=0.1$. We chose a point in parameter space corresponding to $c_s=1$ as this provides more flat UV behaviour in the cross terms which in turn yield interesting phenomenology in the high frequency regime. We sample various $A_\textrm{rel}$ values to show the relative importance of the peak, cross terms and flat contributions of the power and GW spectrum shown in Eqs.~\eqref{eq:PR2} and Eq.~\eqref{eq:split} respectively.

We display our results in Fig.~\ref{fig:crossterms}. The blue curve in all the panels of Fig. \ref{fig:crossterms} corresponds to the log-normal component, the purple corresponds the the cross terms between the step and log-normal component, the cyan corresponds to the step contribution and finally the red curve corresponds to the total. Importantly we see the appearance of the $f^{-2}$ slope for the cross term distinctive of the $c_s^2=1$ \cite{Balaji:2022dbi}. We show the LISA \cite{Barausse:2020rsu}, $\mu$Ares \cite{Sesana:2019vho} and DECIGO \cite{Kawamura:2020pcg} power-law integrated sensitivity curves along with the NANOGrav data.

Observing all the panels of Fig.~\ref{fig:crossterms} we see that for the chosen $k_{\rm p}$, the log-normal peak fits the NANOGrav data well and dominates in the IR tail. However, due to the sharp characteristic cutoff at $f>10^{-7}$ Hz, this component will not be visible to even optimistic proposed GW experiments. However, for $A_\textrm{rel}=10^{-3}$ we note that cross term contribution dominates at frequencies  $10^{-7} \,{\rm  Hz}\lesssim f \lesssim 10^{-6}\,{\rm  Hz}$. The cross term is also sufficiently subdominant to not interfere with the IR behaviour of the total spectrum. At frequencies above $\simeq 10^{-6}$ Hz, the step term dominates and continues as a constant into the high frequency regime.  We also find that the $f^{-2}$ scaling is not enough to reach LISA which is at least $2$ decades away from the peak of the SIGWs needed to explain the PTA data. This means that the signal would be at least a factor $10^{-4}$ suppressed at the scale corresponding to the LISA sensitivity. Since from PTAs we have that $\Omega_{\rm GW}^{\rm peak}\sim 10^{-7}$ at $f\sim 10^{-7}\, \rm Hz$, this takes us to $\Omega^{\rm cross}_{\rm GW}< 10^{-12}$ at $f\sim 10^{-5}\, \rm Hz$, right below LISA. In order to see the $f^{-2}$ scaling we need a detector at $\mu\rm Hz$ such as $\mu$Ares. 

From the right and bottom panels of Fig. \ref{fig:crossterms} we see that for $A_{\rm flat}\gtrsim 2\times10^{-5}$ the plateau would be visible by LISA and DECIGO. Also in that case, we will be able to see the characteristic $f^{-2}$ signature of $c_s^2=1$. 
For $A_{\rm flat}\lesssim 2\times10^{-5}$ the plateau would only be visible to $\mu$Ares and DECIGO. While this is a compelling possibility, we would then not be able to detect the characteristic $f^{-2}$ slope. Nevertheless, we find that for $A_{\rm flat}>10^{-6}$ the plateau would be visible to $\mu$Ares, probing a significantly small amplitude of primordial fluctuations, about three orders of magnitude above the CMB normalisation.

\section{Conclusions\label{sec:conclusions}}

PTA collaborations \cite{NG15-SGWB,NG15-pulsars,EPTA2-SGWB,EPTA2-pulsars,EPTA2-SMBHB-NP,PPTA3-SGWB,PPTA3-pulsars,PPTA3-SMBHB,CPTA-SGWB} reported mounting evidence for a SGWB at nHz frequencies. If such SGWB had cosmic origins, PTAs could be actually probing the physics in a period of the unexplored early universe. Thus, we can use PTAs to test the content of the universe at the time of GW generation, in addition to the physics that lead to the GW generation. 

In this paper, we considered SIGWs produced 
in a early universe dominated by a scalar field and focused on the impact of a different propagation speed of scalar fluctuations while fixing $w=1/3$. 
We find that while PTA data is consistent with $c_s^2=w=1/3$ (see the posterior distributions in Fig.~\ref{fig:fitPTA}), PBH overproduction constraints require $c_s^2>1/3$ (unless specific non-Gaussianities are introduced in the model~\cite{Franciolini:2023pbf}). Remarkably, if we fix $c_s^2=1$, the SIGWs can explain the PTA signal and predict a consistent abundance of PBHs.

Finally, we argued that a distinctive feature of $c_s^2=1$ in our model could be a $\Omega_{\rm GW}\propto f^{-2}$ scaling at the $\mu\rm Hz$ frequencies, which would be probed in the distant future by detectors such as $\mu$Ares. 
We stress that such $f^{-2}$ scaling requires additional power in the curvature fluctuations beyond the log-normal peak potentially associated to the SGWB observed with PTA.
This is present in some inflationary models with sharp features (e.g. see Refs.~\cite{Pi:2017gih,Ando:2018nge,Atal:2018neu,Chen:2019zza,Braglia:2020eai,Ragavendra:2020sop,Fumagalli:2020adf,Ng:2021hll}). In these scenarios, in addition to the $f^{-2}$ tail, we find that in general there should be a flat SIGW spectrum contribution entering the LISA and DECIGO sensitivity (see Fig.~\ref{fig:crossterms}). While the SIGW plateau is also present in other models, e.g. see Ref.~\cite{DeLuca:2020agl,Franciolini:2022pav,Ferrante:2023bgz,Firouzjahi:2023lzg}, its combination with the $f^{-2}$ scaling at $\mu\rm Hz$ may represent a peculiar characteristic of $c_s^2=1$. This could provide an additional way to uncover the content of the universe at the time of wave generation using GWs in PTAs, $\mu$Ares, LISA and DECIGO.

\section*{Acknowledgments} 
We would like to thank A.~Escriv\`a and S.~M.~Young for useful correspondence on PBH formation and Caner Unal for interesting discussions at an early stage of this project.  S.B. is supported by 
funding from the European Union's Horizon 2020 
research and innovation programme under grant 
agreement No 101002846 (ERC CoG ``CosmoChart'') 
as well as support from the Initiative Physique 
des Infinis (IPI), a research training program of the Idex SUPER at Sorbonne
Universit\'{e}.
G.D. is supported by the DFG under the Emmy-Noether program grant no. DO 2574/1-1, project number 496592360. 
G.F. acknowledges the financial support provided under the European Union's H2020 ERC, Starting Grant agreement no.~DarkGRA--757480 and under the MIUR PRIN programme, and support from the Amaldi Research Center funded by the MIUR program ``Dipartimento di Eccellenza" (CUP:~B81I18001170001). This work was supported by the EU Horizon 2020 Research and Innovation Programme under the Marie Sklodowska-Curie Grant Agreement No. 101007855 and additional financial support provided by ``Progetti per Avvio alla Ricerca - Tipo 2", protocol number AR2221816C515921. 
Some calculations of the SIGW spectrum have been performed with  \href{https://github.com/Lukas-T-W/SIGWfast/releases}{\textsc{SIGWfast}} \cite{2022arXiv220905296W}.

\appendix

\section{PBH abundance using peaks and Press-Schecter \label{app:abundance}}

Here we write explicitly the equations and parameters used to calculate the PBH abundance $f_{\rm PBH}$. We follow \cite{Young:2019yug,Gow:2020bzo}. The variance of fluctuations smoothed over a scale $r_m$ is given by\footnote{Notice that uncertainties related to the choice of the window function exists, even though the impact of the choice is reduced when both the variance and the threshold are computed accordingly \cite{Young:2019osy}. }
\begin{align}
\sigma^2  = \frac{16}{81}\int_0^\infty \frac{\mathrm{d}k}{k}(k r_m)^4 {W}^2(kr_m) T^2 (c_sk r_m) \mathcal{P}_{\cal R}(k),
\end{align}
where $W(kr_m)$ is the window function, which for a real space top-hat is given by
\begin{align}
{W}(kr_m) = 3 \frac{\mathrm{sin}(k r_m)- k r_m \mathrm{cos}(k r_m)}{(k r_m)^3},
\end{align}
and $T(c_skr_m)$ is the linear transfer function, which for $w=1/3$ and constant $c_s$ reads
\begin{align}
T(c_skr_m) = 3 \frac{\mathrm{sin}(c_sk r_m)- c_sk r_m \mathrm{cos}(c_sk r_m)}{(c_sk r_m)^3}\,.
\end{align}
For a Dirac delta primordial spectrum it has been shown that $k_{\rm p}r_m\approx 2.744$ \cite{Young:2019osy} while for a log-normal \eqref{eq:PRLN} with $\Delta=0.1$ we have $k_{\rm p}r_m\approx 2.7$ \cite{Musco:2020jjb}.

We will use peaks theory and threshold (Press-Schecter) statistics. For peaks, the number density of peaks is then given by
\begin{align}
{\cal N}(\nu)=\frac{1}{3^{3/2}4\pi^2}\left( \frac{\mu}{\sigma} \right)^3 \nu^3 \exp \left( -\frac{\nu^2}{2} \right)\,,
\end{align}
where $\nu=\delta_l/\sigma$, and
\begin{align}
\mu^2  = \frac{16}{81}\int\limits_0^\infty \frac{\mathrm{d}k}{k}(k r_m)^6 {W}^2(kr_m) T^2 (c_skr_m) \mathcal{P}_{\cal R}(k)\,.
\end{align}

We include unavoidable non-linear effects by relating the linear density fluctuation $\delta_l$ with the non-linear one by
\begin{align}
\delta_m = \delta_l - \frac{3}{8}\delta_l^2\,.
\end{align} 
Taking into account critical collapse, we have that the PBH mass follows
\begin{align}
M_{{\rm PBH}} = \mathcal{K} M_{\rm H} \left( \delta_m - \delta_{\rm th} \right)^\gamma,
\end{align}
with $\mathcal{K}=4$ and $\gamma= 0.36$ as in \cite{Gow:2020bzo}. For $c_s^2=1/3$ we take $\delta_{\rm th}=0.59$ \cite{Musco:2020jjb,Musco:2023dak} and for $c_s^2=1$ we use $\delta_{\rm th}=0.66$. This also fixes the minimum density fluctuation which can give rise to a PBH, namely
\begin{align}
\delta_{c,l-} = \frac{4}{3}\left( 1 - \sqrt{\frac{2-3\delta_{\rm th}}{2}} \right)\,.
\end{align}
With all the above, the total fraction of PBHs at formation reads
\begin{align}
\beta_{\rm peaks} = \int\limits_{\delta_{c,l-}}^{\frac{4}{3}} \mathrm{d}\delta_l \,{\mathcal{K}} \left( \delta_m - \delta_{\rm th} \right)^\gamma {\cal N}(\nu) ,
\end{align}
where the upper integration limit comes from considering only Type-I perturbations (even though this has a very minor impact on results \cite{Musco:2018rwt}).
If we use threshold (Press-Schecter) statistics, we then need to compute \cite{Gow:2020bzo}
\begin{align}
\beta_{\rm PS} = \int\limits_{\delta_{c,l-}}^{\frac{4}{3}} \mathrm{d}\delta_l \,{\mathcal{K}} \left( \delta_m - \delta_{\rm th} \right)^\gamma  \frac{1}{\sqrt{2\pi}\sigma}\exp \left( -\frac{\nu^2}{2} \right)\,.
\end{align}

\section{Scalar field model with general constant \texorpdfstring{$w$}{} and \texorpdfstring{$c_s$}{} \label{app:model}}
In this appendix we provide concrete realizations of model with constant $w$ and $c_s$. We consider a general k-essence type scalar field model \cite{Armendariz-Picon:1999hyi,Garriga:1999vw}, namely the scalar action is given by
\begin{align}
S(\phi)=\int d^4x \sqrt{-g} K(\phi,X)\,,
\end{align}
where $X=-\tfrac{1}{2}\partial_\mu\phi\partial^\mu\phi$. In this model we have that the energy density and pressure are given by
\begin{align}
\rho=2XK_X-K\quad{\rm and}\quad p=K\,,
\end{align}
where $K_X\equiv dK/dX$. The sound speed of fluctuations reads
\begin{align}
c_s^{-2}=1+\frac{2XK_{XX}}{K_X}\,.
\end{align}
The Klein-Gordon equation at the Friedmann–Lemaître–Robertson–Walker (FLRW) background leads to
\begin{align}\label{eq:KG}
\frac{d}{dt}\left(a^3K_X\dot\phi\right)=a^3 K_\phi\,,
\end{align}
where $H=\dot a/a$ with $a$ being the scale factor and $\dot\phi=d\phi/dt$ with $t$ the cosmic time.

The simplest model was proposed by Lucchin and Matarrese \cite{Lucchin:1984yf} and considers
\begin{align}\label{eq:exponential}
K(\phi,X)=X-V_\star e^{\lambda\phi}\,.
\end{align}
This choice leads to $c_s^2=1$ and it is easy to check that it also has a solution of the type $a\propto t^p$, $H=p/t$ and $\phi=2/\lambda\times\ln t$, where $p=2/\lambda^2$. The equation of state is related to $p$ by
\begin{align}\label{eq:wwww}
 w=\frac{2-3p}{3p}\,.
 \end{align}
For $p=1/2$ ($\lambda=2$) we have $w=1/3$.

We can also generalize this model to
\begin{align}
K(\phi,X)=X^\alpha-V_\star \phi^\beta\,.
\end{align}
In this case one finds that
\begin{align}
c_s^{-2}=2\alpha-1\,.
\end{align}
Furthermore, requiring a solution of the type $a=a_\star(t/t_\star)^p$ and $\phi=\phi_\star(t/t_\star)^q$ gives a relation between $\alpha$ and $\beta$. They must also must satisfy the Friedmann equation, namely $3H^2=\rho$. Equating the powers of $t$ in Eq.~\eqref{eq:KG} and in the Friedmann equation, we find
\begin{align}
q=\frac{\alpha-1}{\alpha}\quad {\rm and}\quad \beta=\frac{2\alpha}{1-\alpha}\,.
\end{align}
The equation of state for this model is again given by Eq.~\eqref{eq:wwww}. Solving the Klein-Gordon and Friedmann equation relates other parameters such as $V_\star$ and $\phi_\star$ with $\alpha$ and $p$.  In this way, we have a model with constant and independent $w$ and $c_s$. As an example, when $V_\star=0$ we have $c_s^2=w$. We can also formally recover the $c_s^2=1$ case in the limit where $q\to 0$ while keeping $\phi_\star\times q={\rm constant}$. Solving the Klein-Gordon and Friedmann equations yields $q\phi_\star=\sqrt{2p}$, consistent with the parameters of the exponential potential (see the discussion after Eq.~\eqref{eq:exponential}).

\section{Additional results with fixed \texorpdfstring{$c_s^2$}{ } \label{app:fixedcs}}

\begin{figure}[!htp]
\includegraphics[width=0.49\columnwidth]{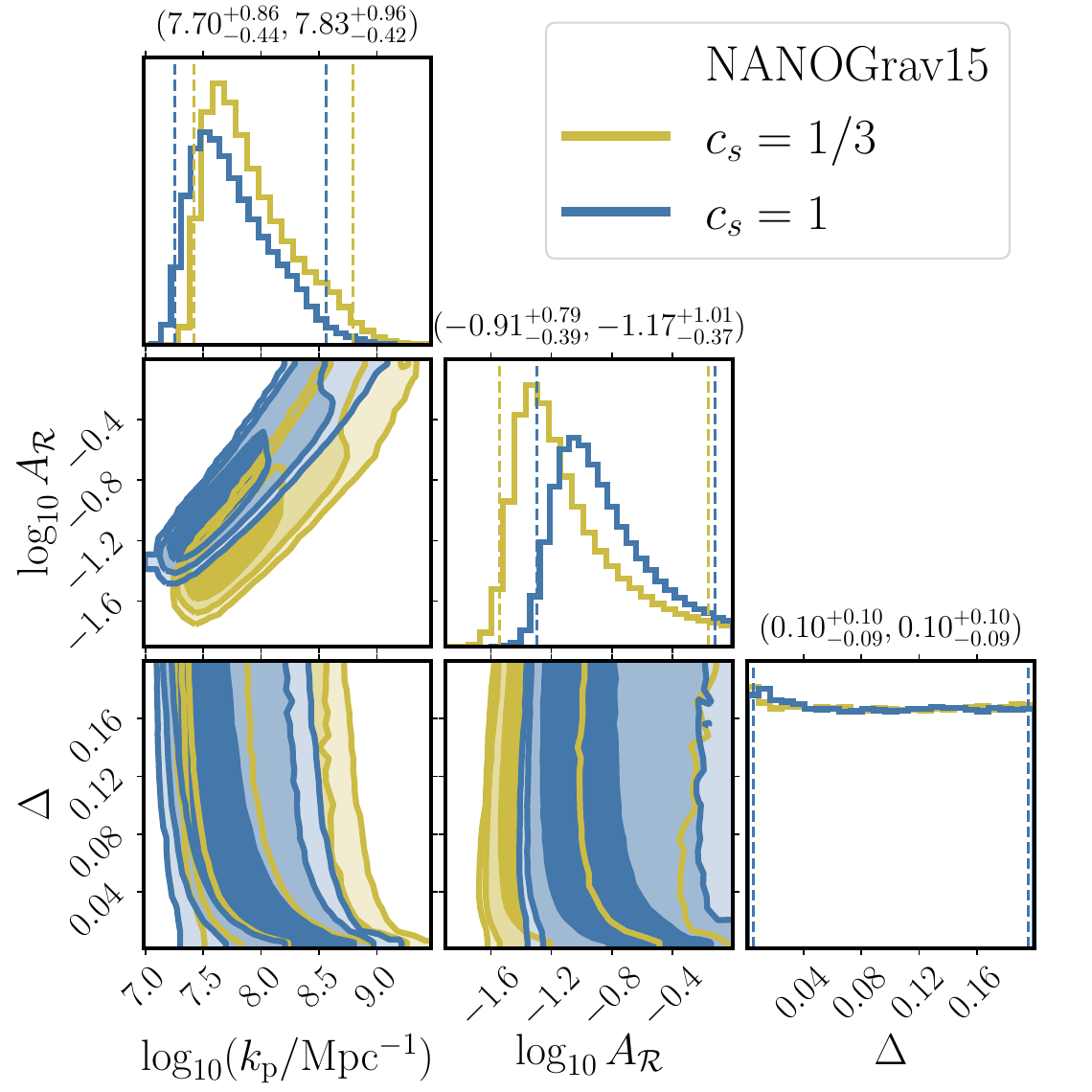}
\includegraphics[width=0.49\columnwidth]{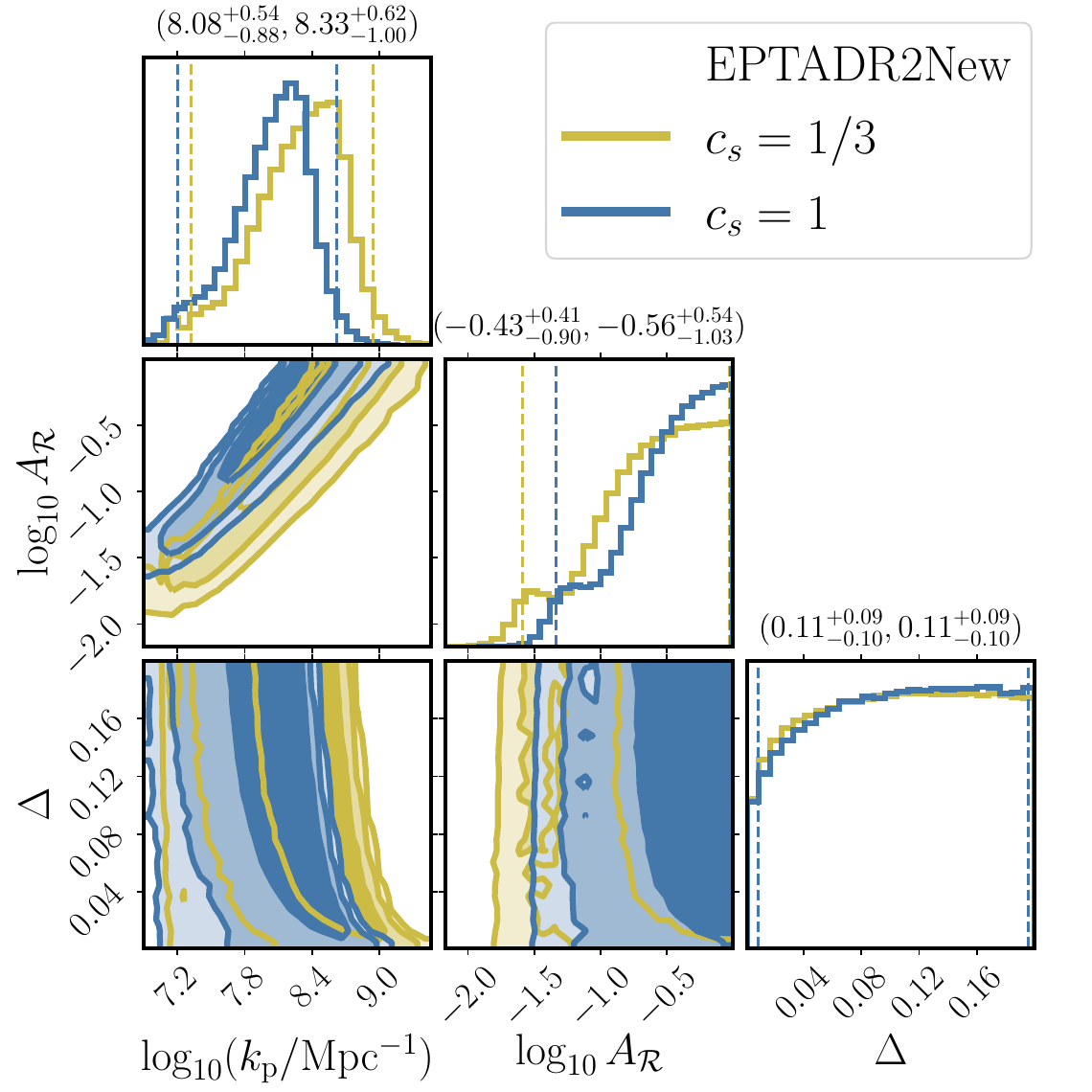}
\caption{ Posterior distributions for NANOGrav (left panel) and EPTA (right panel) for SIGWs generated during a phase with $w=1/3$ and fixed $c_s$. We show the results for $c_s^2=1/3$ in yellow and $c_s^2=1$ in blue. Note how a higher $c_s$ requires higher values of $A_{\cal R}$ and lower values of $k_{\rm p}$ consistent with Eq.~\eqref{eq:relations}.
\label{fig:fitPTAcomparisons}}
\end{figure}

\begin{figure}[!ht]
\includegraphics[width=0.49\columnwidth]{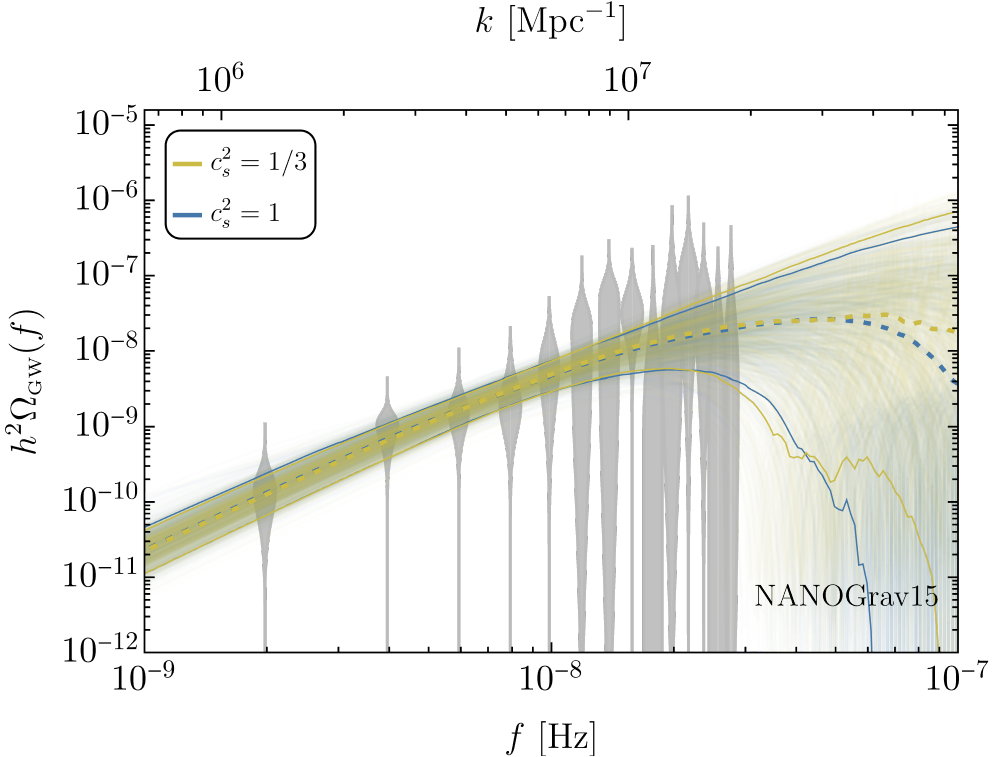}
\includegraphics[width=0.49\columnwidth]{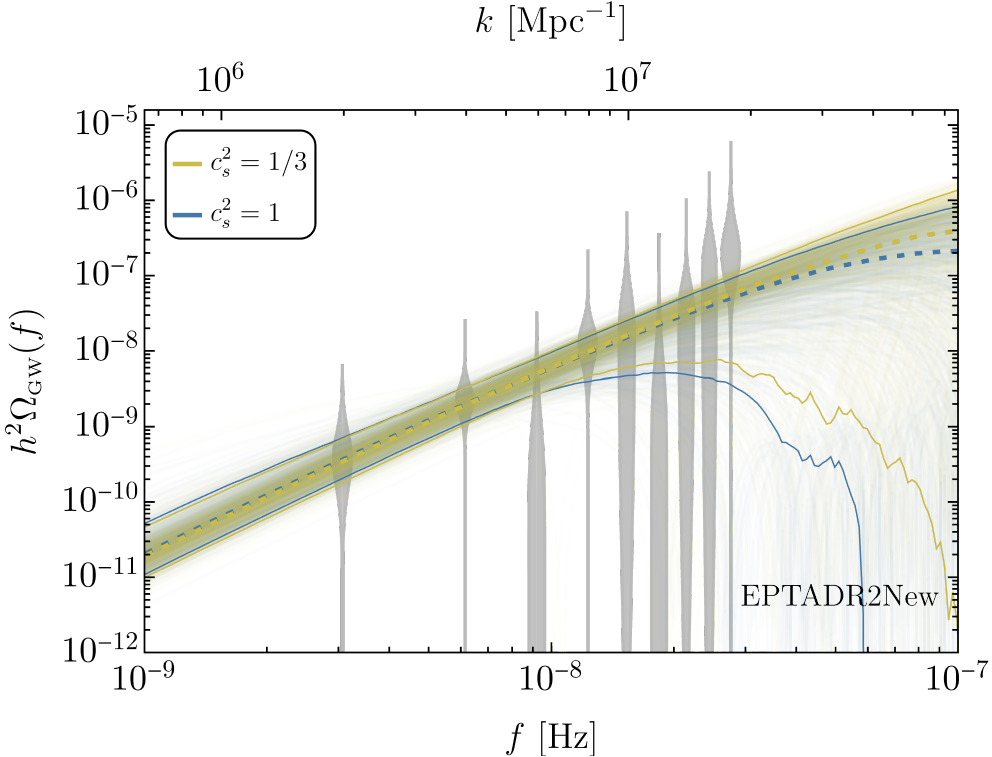}
\caption{  SIGW spectral density for fixed $c_s$ parameter (see Eqs.~\eqref{eq:OmegaDelta} and \eqref{eq:spectraldensitytodaydirac}). We respectively show the fit to NANOGrav15 and EPTA data sets on the left and right panels. The colored bands show the 90\% credibility intervals and the gray violins show the posteriors from NANOGrav \cite{NG15-pulsars,NG15-NP} and EPTA \cite{EPTA2-pulsars}.
\label{fig:spectraapp}}
\end{figure}

In this appendix we provide the results of the analysis for fixed $c_s^2=1/3$ and $c_s^2=1$. We take the priors shown in Tab.~\ref{tab:priors}. 
We show the resulting posteriors in Fig.~\ref{fig:fitPTAcomparisons} and the violin plots with the SIGW spectrum in Fig.~\ref{fig:spectraapp}. We find that a higher $c_s$ requires a higher amplitude $A_{\cal R}$ of the primordial spectrum and a lower value of $k_{\rm p}$. As we show in Fig.~\ref{fig:pbhresults}, the increase in $c_s^2$ has a bigger impact on $f_{\rm PBH}$ than the increase of $A_{\cal R}$ leading to a fraction of PBHs consistent with the $1\sigma$ bounds from the PTA analysis. We also see from Fig.~\ref{fig:spectraapp} that the data is mainly fitted by the low frequency tail of the SIGW spectrum.

\begin{table}[!ht]
\begin{tabularx}{0.7\textwidth}{|| l || Y Y Y Y||} 
 \hline
 \textbf{Parameter} & $\log_{10}A_{\cal R} $& $\log_{10}(k_{\rm p}\,[\rm Mpc^{-1}])$ & $c_s$ & $\Delta$ \\ [0.5ex] 
 \hline\hline
 \textbf{Prior} & $[-10:0]$ & $[5:10]$ & $[0:1] $& $[0:0.2]$ \\ 
 \hline
\end{tabularx}
\caption{Uniform priors for the Bayesian analysis for free $c_s$. We also use the same priors on the remaining parameters in the analyses that fixes $c_s$ to specific values. \label{tab:priors}}
\end{table}

\bibliography{refgwscalar.bib} 

\end{document}